\documentclass[twocolumn,english,aps,prb,twocolumn,10pt,amsfonts,amssymb,amsmath,longbibliography,floatfix,superscriptaddress,nofootinbib]{revtex4-2}

\usepackage{dcolumn}
\usepackage[T1]{fontenc}
\usepackage[utf8]{inputenc}
\usepackage[dvipsnames]{xcolor}
\usepackage{babel}
\usepackage{array}
\usepackage{float}
\usepackage{bm}
\usepackage{multirow}
\usepackage{amsmath}
\usepackage{amssymb}
\usepackage{graphicx}
\usepackage[normalem]{ulem}
\usepackage{soul}
\usepackage{xspace}
\usepackage{hyperref}
\hypersetup{
    unicode=true,
    bookmarksnumbered=false,
    bookmarksopen=false,
    breaklinks=true,
    pdfborder={0 0 0},
    pdfborderstyle={},
    colorlinks,
    citecolor=blue,
    filecolor=blue,
    linkcolor=blue,
    urlcolor=blue
}

\makeatletter

\floatstyle{ruled}
\newfloat{algorithm}{H}{loa}
\providecommand{\algorithmname}{Algorithm}
\floatname{algorithm}{\protect\algorithmname}

\usepackage{physics}
\providecommand{\kp}{\ensuremath{\bm{k}\cdot\bm{p}}\xspace}
\definecolor{LightGray}{gray}{0.9}

\makeatother

\usepackage[frozencache,cachedir=Minted]{minted}
\setminted{mathescape=true,
bgcolor=LightGray,
numberblanklines=false,
python3=true,
numbersep=2pt,
samepage=true,
escapeinside={||}}

\begin{document}

\title{
\texorpdfstring{
    \texttt{DFT2kp}: effective $\mathbf{k}\cdot\mathbf{p}$ models from \emph{ab initio} data
}{
    DFT2kp: effective kp models from ab initio data
}
}

\author{João Victor V. Cassiano}
\affiliation{Instituto de Física, Universidade Federal de Uberlândia, Uberlândia, MG 38400-902, Brazil}
\affiliation{Instituto de Física, Universidade de São Paulo, São Paulo, SP, 05508-090, Brazil}

\author{Augusto L. Ara\'ujo}
\affiliation{Instituto de Física, Universidade Federal de Uberlândia, Uberlândia, MG 38400-902, Brazil}
\affiliation{Ilum School of Science, CNPEM, C.P. 6192, 13083-970, Campinas, SP, Brazil}

\author{Paulo E. {Faria~Junior}}
\affiliation{Institute of Theoretical Physics, University of Regensburg, 93040 Regensburg, Germany}

\author{Gerson J. Ferreira}
\affiliation{Instituto de Física, Universidade Federal de Uberlândia, Uberlândia, MG 38400-902, Brazil}

\begin{abstract}
The $\bm{k}\cdot\bm{p}$ method, combined with
group theory, is an efficient approach to obtain
the low energy effective Hamiltonians of crystalline materials. Although the Hamiltonian coefficients are written as matrix elements of the generalized momentum operator $\bm{\pi}=\bm{p}+\bm{p}_{{\rm SOC}}$ (including spin-orbit coupling corrections), their numerical values must be determined from outside sources, such as experiments or \emph{ab initio} methods.
Here, we develop a code to explicitly calculate the Kane (linear in crystal momentum) and Luttinger (quadratic in crystal momentum) parameters of $\bm{k}\cdot\bm{p}$ effective Hamiltonians directly from \emph{ab initio} wavefunctions provided by Quantum ESPRESSO. Additionally, the code analyzes the symmetry transformations of the wavefunctions to optimize the final Hamiltonian. This is an optional step in the code, where it numerically finds the unitary transformation $U$ that rotates the basis towards an optimal symmetry-adapted representation informed by the user. Throughout the paper, we present the methodology in detail and illustrate the capabilities of the code applying it to a selection of relevant materials. Particularly, we show a ``hands-on'' example of how to run the code for graphene (with and without spin-orbit coupling). The code is open source and available at \href{https://gitlab.com/dft2kp/dft2kp}{https://gitlab.com/dft2kp/dft2kp}.
\end{abstract}
\maketitle

\section{Introduction}

The band structure of crystalline materials defines most of its electronic properties, and its accurate description is essential to the development of novel devices. For this reason, the \emph{ab initio} density functional theory (DFT) \cite{dft1,dft2} provides one of the most successful tools for the development of electronics, spintronics, optoelectronics, etc. The DFT methods have been implemented in a series of codes (\textit{e.g.}, Quantum ESPRESSO \cite{Giannozzi2009, Giannozzi2017}, VASP \cite{VASP}, Wien2K \cite{Wien2k}, Gaussian \cite{g16}, DFTB$+$ \cite{DFTBplus}, Siesta \cite{Siesta1, Siesta2}, ...), which differ by the choice of basis functions (\textit{e.g.}, localized orbitals or plane-waves), pseudo-potential approximations, and other functionalities. Nevertheless, all DFT implementations provide methods to obtain the equilibrium (relaxed) crystalline structure, phonon dispersion, and electronic band structures. Complementary, few bands effective models are essential to further study transport, optical, and magnetic properties of crystalline materials. These can be developed either via the tight-binding (TB) \cite{SlaterKoster, Goringe1997, yu2005fundamentals} or \kp method \cite{Voon2009, Winkler2003}, which complement each other.

On the one hand, the TB method has an ``atomistic'' nature, since it is built upon localized basis sets (\textit{e.g.}, maximally-localized Wannier functions \cite{Marzari2012Wannier}, or atomic orbitals), which makes this method optimal for numerical modeling of transport, optical and other properties of complex \textit{nanomaterials} \cite{Persson2004PRB, Soluyanov2016PRB,Ridolfi2017PRB, Frank2018PRL}.

On the other hand, the \kp method uses basis sets of extended waves, which are exact solutions of the Hamiltonian at a quasi-momentum of interest, typically at a high symmetry point of the Brillouin zone. While this characteristic may limit the \kp description to a narrow region of the energy-momentum space, 
the \kp Hamiltonians are easier to handle analytically and, especially, are very suitable to study mesoscopic systems using the envelope function approximation \cite{Bastard1981:PRB,
Burt1988a:SST,Burt1988b:SST,Baraff1991:PRB,Burt1992:JPCM,Foreman1996:PRB}. For example, the k.p framework has been successfully applied to study nanostructures (quantum wells, wires, and dots) \cite{Pryor2006PRL, Campos2018,vanBree2012PRB}, topological insulators \cite{Novik2005:PRB, Bernevig2006,Miao2012:PRL}, spin-lasers \cite{Holub2011PRB, FariaJunior2015PRB}, polytypism \cite{FariaJunior2012JAP, FariaJunior2014JAP,Climente2016:JAP}, as well as a large variety of two-dimensional van der Waals materials \cite{Li2014:PRB, Kormanyos2015,Li2015PRB, FariaJunior2019PRB}. Moreover, recent developments in the field of transition metal dichalcogenides (TMDCs) have combined DFT and \kp methodologies to explore the valley Zeeman physics in TMDC monolayers and their van der Waals heterostructures \cite{Wozniak2020PRB, Deilmann2020PRL, Forste2020NatComm, Xuan2020PRR}.

Both the TB and \kp Hamiltonians are defined in terms of arbitrary coefficients. In the TB case, these are local site energies and hopping amplitudes described by Slater-Koster matrix elements \cite{SlaterKoster}. For the \kp Hamiltonians, these are the Kane \cite{Kane1956, Kane1957} and Luttinger \cite{Luttinger1955} parameters, which are matrix elements of the momentum and spin-orbit coupling operators. 
In both methods (TB or \kp), the values of these arbitrary coefficients must be determined from outside sources, which strongly depend on the size and analytical properties of the particular model Hamiltonian. For instance, early studies within the \kp framework have shown that for parabolic single band descriptions, or weakly coupled models, it is possible to write the quadratic coefficients in terms of effective masses, which can be experimentally determined by cyclotron resonance experiments \cite{Dexter1954PR, Dresselhaus1955PR, Kane1956, Kane1959JPCS, Wallis1960PR}. Moreover, energy splittings, such as band gaps, can be directly determined from optical experiments \cite{Cardona1967PR, Aspnes1973PRB, Gadret2010PRB, DeLuca2015NL, Zilli2015ACSNano}. For III-V semiconductors with zinc-blend structure and nitride-based wurtzite compounds, a useful database for \kp parameters inspired by experimentally available datasets can be found in Ref.~\cite{Vurgaftman2001JAP}. Conversely, for \kp Hamiltonians that do not allow analytical solutions, but still have a low number of bands ($\sim 10$), it is possible to perform numerical fitting techniques to DFT calculations \cite{Kormanyos2015, Bastos2016SST, Paulo2016, Winkler2016PRL, Nechaev2016PRB, Bastos2018JAP, Zhou2017PRB, FariaJunior2019PRB, Pulcu2023arXiv}. For larger \kp Hamiltonians ($>30$ bands), fitting procedures may also be applied \cite{Beresford2004, Rideau2006PRB} or directly extracted from first principles calculations, since the only matrix elements involved are linear in momentum \cite{Persson2007, Shishidou2008, Berland2017}. Interestingly, these large band \kp models can even be used to supplement and speed up first principles calculations, as demonstrated in Refs.~\cite{Persson2007, Shishidou2008, Berland2017}. 
In TB models, fitting procedures can also be applied to obtain the unknown parameters \cite{Konschuh2010PRB, Cappelluti2013PRB, Ridolfi2015JPCM, Gmitra2016PRB, Menezes2018CMS}. 
Conversely, fully automated procedures, integrated within \emph{ab initio} codes, such as the \texttt{wannier90} code \cite{Mostofi2008, Mostofi2014}, use localized Wannier functions computed from the DFT wave functions to calculate TB parameters. Moreover, explicit calculations of the Slater-Koster matrix elements are implemented in the \texttt{paoflow} \cite{BuongiornoNardelli2018} and DFTB+ \cite{DFTBplus} codes.

While it is possible to extract \kp models from a Taylor expansion on top of a TB model (\textit{e.g.,} via the code \texttt{tbmodels} \cite{Gresch2018}), there are no versatile implementations to calculate the \kp Kane (linear in k) and Luttinger (quadratic in $\bm{k}$) parameters directly from the DFT wavefunctions \cite{VASP2kpNote}. 
To calculate the \kp matrix elements from the DFT wavefunctions, one needs to account for how the wavefunctions are represented in the DFT code \cite{Pickard2000, Shishidou2008}.
For instance, Quantum ESPRESSO and VASP implement pseudopotential approximations within the Projector Augmented Wave (PAW) method \cite{Bloechl1994, Kresse1999, Corso2010, Schwerdtfeger2011}. 
Fortunately, Quantum ESPRESSO already provides a routine to calculate matrix elements of the velocity operator (which is sufficient to obtain \kp models, as we see in Section \ref{sec:DFT}). Indeed, recently, Joci\'c and collaborators \cite{Jocic2020} have successfully calculated \kp models directly from QE's wavefunctions (see disclaimer at our Conclusions).

In this paper, we present an \textit{open-source code} that automatically calculates the numerical values for the \kp Kane and Luttinger parameters using the wavefunctions provided by Quantum ESPRESSO (QE). For this purpose, first, we develop a patch to instruct QE to calculate and store the matrix elements of the generalized momentum $\bm{\pi} = \bm{p} + \bm{p}_{\rm SOC}$, which includes the spin-orbit corrections. Together with the eigenenergies $E_n^0$ at $\bm{k}_0$, the matrix elements of $\bm{\pi}$ for a selected set of $N$ bands define the effective \kp Hamiltonian $H_{N\times N}(\bm{k})$ for $\bm{k}$ near $\bm{k}_0$. Our \texttt{python} package reads these matrix elements and QE's wavefunctions $\ket{n}$ to automatically build $H_{N\times N}(\bm{k})$ using L\"owdin's partitioning \cite{Lowdin1951} for the folding down of all QE bands into the selected $N$ bands subspace. Additionally, the user has the option to improve the appearance (or \emph{form}) of the effective Hamiltonian via a symmetry optimization process aided by the \texttt{qsymm} package \cite{Varjas2018}, which builds the symbolic Hamiltonian via group theory and the method of invariants. To illustrate the capabilities of our code, we show here a step-by-step ``hands-on'' tutorial on how to run the code for graphene, and later we present results for selected materials [zincblende, wurtzite, rock-salt, transition metal dichalcogenides (TMDC), and others]. In all cases, the modeled band structure matches remarkably well the DFT data at low energies near the expansion point $\bm{k}_0$. Our code is open source and available at the \texttt{gitlab} repository \cite{DFT2KP}.

This paper is organized as follows. In Section \ref{sec:Methods} we present our methodology starting with a brief review of the \kp method, L\"owdin partitioning, the method of invariants, the symmetry optimization process, and the calculation of matrix elements using the DFT data. Next, in Section \ref{sec:graphene}, we show the code in detail using graphene as a practical example. Later, in Section \ref{sec:Examples}, we illustrate the results of the code for zincblend (GaAs, CdTe, HgTe), wurtzite (GaP, GaN, InP), rock-salt (SnTe, PbSe), a TMDC (${\rm MoS}_2$), and other materials (${\rm Bi}_2{\rm Se}_3$, ${\rm GaBiCl}_2$). We finish the paper with an overview of the results in Section \ref{sec:discussions}, and the conclusions.

\section{Methods}
\label{sec:Methods}

Our goal is to obtain the numerical values for the coefficients of
$\bm{k}\cdot\bm{p}$ effective Hamiltonians \cite{Winkler2003,Voon2009}.
Namely, these are the Kane \cite{Kane1956,Kane1957} and Luttinger
\cite{Luttinger1955} parameters. To present our approach to this
calculation, let us start by briefly describing its fundamental steps.
First, we review the $\bm{k}\cdot\bm{p}$ method to show that these
coefficients depend only upon matrix elements of the type $\bm{P}_{m,n}=\bra{m}\bm{\pi}\ket{n}$,
where $\bm{\pi}=\bm{p}+\bm{p}_{{\rm SOC}}$ is the generalized momentum
operator with the spin-orbit corrections, and $\{\ket{n}\}$ is the
set of numerical wavefunctions obtained from the \emph{ab initio} DFT simulations (e.g., via Quantum ESPRESSO \cite{Giannozzi2009,Giannozzi2017}).
However, the numerical DFT basis given by $\{\ket{n}\}$ does not match,
\emph{a priori}, the optimal symmetry-adapted basis set that yields the desired form
for the effective $\bm{k}\cdot\bm{p}$ Hamiltonian. Therefore, to properly
identify the Kane and Luttinger parameters, we perform a \emph{symmetry
optimization}, which rotates the arbitrary numerical basis into the
optimal symmetry-adapted form. This symmetry optimization is performed via group theory
\cite{Tinkham2003,Dresselhaus2007} by enforcing that the numerical
DFT basis transforms under the same representation of an \emph{optimal symmetry-adapted basis}, which is informed by the user. \newline

\noindent In summary, the algorithm steps are:

\begin{enumerate}
\item Read the QE/DFT data: energies $E_{n}^{0}$ and eigenstates $\ket{n}$
at the selected $\bm{k}_0$ point.
\item Calculate or read the matrix elements of $\bm{P}_{m,n}=\bra{m}\bm{\pi}\ket{n}$
for all bands $(m,n)$. 
\item Select the bands of interest (set $A$). The code will identify
the irreducible representations of the bands using the \texttt{IrRep}
python package \cite{IrRep}, and present it as a report to the user.
Additionally, the code calculates the model folded down into the selected
set $A$ via L\"owdin partitioning.
\item Build the optimal effective model from symmetry constraints using
the \texttt{Qsymm} python package \cite{Varjas2018} under an optimal symmetry-adapted
basis informed by the user. This optimal basis must be in a set of
representations equivalent to the ones identified in Step 3.
\item Calculate the representation matrices for the symmetry operators in
the original QE basis $\ket{n}$. The code verifies if the representations
of the numerical QE basis are equivalent to the representations of the
optimal symmetry-adapted basis from step 4.
\item Calculates the transformation matrix $U$ that rotates the original
QE basis into the optimal symmetry-adapted basis set in step 4. Applies the transformation
$U$ and calculates the optimal symmetry-adapted numerical effective Hamiltonian.
\item Convert values from Rydberg atomic units into meV and nm units, and
present a report with values for the $\bm{k}\cdot\bm{p}$ parameters.
\end{enumerate}

In the next sections, we describe the relevant details of the steps
above, but not following the algorithmic order above. More specifically,
in Section \ref{subsec:kp}, we briefly review the $\bm{k}\cdot\bm{p}$
formalism to show that $\bm{P}_{m,n}=\bra{m}\bm{\pi}\ket{n}$ plays
a central role in our approach. Incidentally, we introduce the folding
down via L\"owdin partitioning \cite{Lowdin1951}. Next, we define what is the optimal symmetry-adapted
form of the Hamiltonian via the method of invariants \cite{Luttinger1956,Winkler2003}
in Section \ref{subsec:optimal}. In Section \ref{subsec:symmopt},
we present the symmetry optimization approach to calculate the transformation
matrix $U$ that yields our final $H^{{\rm optimal}}=U\cdot H^{{\rm DFT}}\cdot U^{\dagger}$.
At last, in Section \ref{subsec:matelDFT} we discuss how $\bm{P}_{m,n}=\bra{m}\bm{\pi}\ket{n}$
is calculated.

Throughout the paper we use atomic Rydberg units (a.u.), thus the
reduced Planck constant, bare electron mass and charge are $\hbar=2m_{0}=e^{2}/2=1$,
the permittivity of vacuum is $4\pi\varepsilon_{0}=1$, the speed
of light is $c=2/\alpha\approx274$, and $\alpha\approx1/137$ is
the fine structure constant.

\subsection{
\texorpdfstring{
    The $\mathbf{k}\cdot\mathbf{p}$ model
    \label{subsec:kp}
}{
    The kp model
}}

In this section, we briefly review the $\bm{k}\cdot\bm{p}$ method
\cite{Kane1956,Kane1957,Luttinger1955,Winkler2003,Voon2009} and
the folding down via L\"owdin partitioning \cite{Lowdin1951, Luttinger1956, Winkler2003}
to establish our notation. 

We are interested in the effective Hamiltonian near a high-symmetry
point $\bm{k}_{0}$ of the Brillouin zone. Therefore, we write the
quasi-momentum as $\bm{\kappa}=\bm{k}_{0}+\bm{k}$, such that $\bm{k}$
is the deviation from $\bm{k}_{0}$. The Bloch theorem allow us to
decompose the wavefunction as $\psi_{\bm{\kappa}}(\bm{r})=e^{i\bm{k}\cdot\bm{r}}\phi_{\bm{k}_{0},\bm{k}}(\bm{r})$,
with $\phi_{\bm{k}_{0},\bm{k}}(\bm{r})=e^{i\bm{k}_{0}\cdot\bm{r}}u_{\bm{k}_{0}+\bm{k}}(\bm{r})$,
where $u_{\bm{k}_{0}+\bm{k}}(\bm{r})\equiv u_{\bm{\kappa}}(\bm{r})$
is the periodic part of the Bloch function, while $\phi_{\bm{k}_{0},\bm{k}}(\bm{r})$
carries the phase given by $\bm{k}_{0}$ and obeys the Schr\"odinger
equation $[H^{0}+H'(\bm{k})]\phi_{\bm{k}_{0},\bm{k}}(\bm{r})=[E-k^{2}]\phi_{\bm{k}_{0},\bm{k}}(\bm{r})$,
with
\begin{align}
    H^{0} & =p^{2}+V(\bm{r})+2\bm{k}_{0}\cdot\bm{\pi}+H_{{\rm SR}},
    \\
    H'(\bm{k}) & =2\bm{k}\cdot\bm{\pi},
    \\
    \bm{\pi} & =\bm{p}+\frac{\alpha^{2}}{8}\bm{\sigma}\times\nabla V(\bm{r}),
\end{align}
where $H^{0}$ is the Hamiltonian at $\bm{k}=0$,
$V(\bm{r})$ is the periodic potential,
$H'(\bm{k})$ carries
the k-dependent contributions that will be considered as a perturbation
hereafter, $\bm{\pi}$ is the generalized momentum that includes
the spin-orbit contributions (SOC), 
and $\bm{\sigma}=(\sigma_{x},\sigma_{y},\sigma_{z})$ are the Pauli matrices for the electron spin. 
For simplicity, we consider only
leading order corrections of the fine structure terms. Namely, at
$\bm{k}=0$, the $H_{{\rm SR}}$ carries the scalar relativistic terms,
composed by the Darwin, $H_{{\rm D}}=\frac{\alpha^{2}}{8}\nabla^{2}V(\bm{r})$,
and the mass-velocity corrections, $H_{{\rm MV}}=-\alpha^{2}p^{4}/4$.
In the \emph{ab initio} DFT data, these are implied in the numerical
eigenvalues $E_{n}^{0}$ of $H^{0}$. For finite $\bm{k}\neq0$, we
keep only the SOC contribution in $\bm{\pi}$, and neglect the higher
order mass-velocity corrections (see Appendix \ref{sec:massvelocity}).

The DFT data, as shown in the next section, provide us with a set
$\{\ket{n}\}$ of eigenstates of $H^{0}$, i.e. $H^{0}\ket{n}=E_{n}^{0}\ket{n}$.
From this \emph{crude DFT basis}, we define an all bands model $H_{{\rm all}}^{{\rm DFT}}(\bm{k})$,
with matrix elements
\begin{equation}
    \bra{m} H_{{\rm all}}^{{\rm DFT}} \ket{n}    = E_{n}^{0}\delta_{m,n}+2\bm{k}\cdot\bm{P}_{m,n},
    \label{eq:crude_all_bands}
\end{equation}
where $\bm{P}_{m,n}=\bra{m}\bm{\pi}\ket{n}$. We refer to this as
the \emph{crude model} because it is calculated from the original
numerical DFT wavefunctions, which do not have an optimal symmetry-adapted form (more detail in
Section \ref{subsec:symmopt}). Nevertheless, it already shows that
$E_{n}^{0}$ and $\bm{P}_{m,n}$ are central quantities, and both
can be extracted from DFT simulations, as shown in Section \ref{subsec:matelDFT}. 

Next, we want to fold down $H_{{\rm all}}^{{\rm DFT}}$ into a subspace
of $N$ bands near the Fermi energy to obtain our reduced, but still
crude, effective model $H_{N\times N}^{{\rm DFT}}$. This is
done via L\"owdin partitioning \cite{Lowdin1951, Luttinger1956, Winkler2003}. First,
the user must inform the set of $N$ bands of interest, which we refer
to as set $A$. Complementary, the remaining remote bands compose
the set $B$. Considering the diagonal basis $H^{0}\ket{n}=E_{n}^{0}\ket{n}$,
and the perturbation $H'(\bm{k})$, the L\"owdin partitioning leads
to the effective Hamiltonian $H_{N\times N}^{{\rm DFT}}$ defined
by the expansion
\begin{multline}
    [H_{N\times N}^{{\rm DFT}}]_{m,n}(\bm{k})=\Big(E_{n}^{0}+k^{2}\Big)\delta_{m,n}+H'_{m,n}(\bm{k})
    \\
    +\frac{1}{2}\sum_{r\in B}H'_{m,r}(\bm{k})H'_{r,n}(\bm{k})\Bigg(\frac{1}{E_{m}^{0}-E_{r}^{0}}+\frac{1}{E_{n}^{0}-E_{r}^{0}}\Bigg)+\cdots
    \label{eq:lowdin}
\end{multline}
with $H'_{m,n}(\bm{k})=\bra{m}H'(\bm{k})\ket{n}=2\bm{k}\cdot\bm{P}_{m,n}$.
Here, the indices $m,n\in A$ run over the bands we want to model
(set $A$), while $r\in B$ run over the remote bands. The
expansion above is shown up to second order in $H'$, but higher order
terms can be found in Ref. \cite{Winkler2003}. 
Alternatively, the recent python package \texttt{pymablock} \cite{Pymablock} implements an efficient numerical method
to compute the Löwdin partitioning to arbitrary order.

\subsection{The optimal symmetry-adapted form of H\label{subsec:optimal}}

The selection rules from group theory allow us to identify which matrix
elements of an effective Hamiltonian are finite \cite{Tinkham2003}.
More interestingly, the method of invariants \cite{Luttinger1956,Winkler2003}
can be used to directly obtain the most general form of $H_{N\times N}^{{\rm optimal}}(\bm{k})$
allowed by symmetry. To define this form, consider a Taylor
series expansion
\begin{equation}
    H_{N\times N}^{{\rm optimal}}(\bm{k})=\sum_{i,j,l}h_{i,j,l}\;k_{x}^{i}\;k_{y}^{j}\;k_{z}^{l},
\end{equation}
where $h_{i,j,l}$ are constant matrices that multiply the powers
of $\bm{k}=(k_{x},k_{y},k_{z})$ as indicated by its indices 
$i,j,l=\{0,1,2,\dots\}$.
To find the symmetry allowed $h_{i,j,l}$, we recall that the space
group $\mathcal{G}$ of the crystal is defined by symmetry operations
that keep the crystalline structure invariant. Particularly, at a
high symmetry point $\bm{\kappa}=\bm{k}_{0}$, one must consider the
little group $\mathcal{G}_{\bm{k}_{0}}\in\mathcal{G}$ of symmetry
operations that maintain $\bm{k}_{0}$ invariant (the star of $\bm{k}_{0}$).
Hence, $H_{N\times N}^{{\rm optimal}}(\bm{k})$ must commute with
the symmetry operations of $\mathcal{G}_{\bm{k}_{0}}$. Namely,
\begin{equation}
    H_{N\times N}^{{\rm optimal}}(D^{k}(S)\bm{k})=D^{\psi}(S)H_{N\times N}^{{\rm optimal}}(\bm{k})D^{\psi}(S^{-1}),
    \label{eq:invariants}
\end{equation}
where $D^{\psi}(S)$ are the representation matrices for each symmetry
operator $S\in\mathcal{G}_{\bm{k}_{0}}$ in the subspace defined by
the wavefunctions of set $A$, and $D^{k}(S)$ are the representation
matrices acting on the vector $\bm{k}=(k_{x},k_{y},k_{z})$. The set
of equations defined by this relation for all $S\in\mathcal{G}_{\bm{k}_{0}}$
leads to a linear system of equations that constrain the symmetry
allowed form of $H_{N\times N}^{{\rm optimal}}(\bm{k})$, i.e., it
defines which of constant matrices $h_{i,j,l}$ are allowed up to
a multiplicative factor. Ultimately, these multiplicative factors
are the Kane and Luttinger parameters that we want to calculate numerically.

The python package \texttt{Qsymm} \cite{Varjas2018} implements an
efficient algorithm to find the form of $H_{N\times N}^{{\rm optimal}}(\bm{k})$
solving the equation above and returns the symmetry allowed $h_{i,j,l}$.
\texttt{Qsymm} refers to these as the \emph{Hamiltonian family}. To
perform the calculation, the user must inform the representation matrices
$D^{\psi}(S)$ for the generators of $\mathcal{G}_{\bm{k}_{0}}$.
Notice that the choice of representation is arbitrary, and different
choices lead to effective Hamiltonians with different forms.
This ambiguity is the reason the next step, symmetry optimization,
is necessary.

\subsection{Symmetry optimization\label{subsec:symmopt}}

In the previous section, the matrix representations for generators
$S\in\mathcal{G}_{\bm{k}_{0}}$ are implicitly written in an \emph{optimal
symmetry-adapted basis}, which we will now label with an $\mathcal{O}$ index, as in $\{\ket{n_{\mathcal{O}}}\}$,
to distinguish from the \emph{crude DFT} numerical basis, which we now
label with an $\mathcal{C}$ index, as in $\{\ket{n_{\mathcal{C}}}\}$. The matrix representations
of $S$ written in these two bases are equivalent up to a unitary transformation $U$, i.e. $D^{\mathcal{O}}(S)=U\cdot D^{\mathcal{C}}(S)\cdot U^{\dagger}$.
Indeed, this same matrix $U$ transforms the \emph{crude DFT} numerical
Hamiltonian into the desired \emph{optimal symmetry-adapted form}, i.e. $H_{N\times N}^{{\rm optimal}}=U\cdot H_{N\times N}^{{\rm DFT}}\cdot U^{\dagger}$. Therefore, our goal here is to find this transformation matrix $U$.

For each symmetry operator $S_{i}\in\mathcal{G}_{\bm{k}_{0}}$, let
us define $\mathcal{C}^{i}\equiv D^{\mathcal{C}}(S_{i})$ and $\mathcal{O}^{i}\equiv D^{\mathcal{O}}(S_{i})$
as the representation matrices under the original numerical DFT basis
($\mathcal{C}$), and under the desired optimal symmetry-adapted representation ($\mathcal{O}$), respectively.
For irreducible representations, this $U$ is unique (modulo a phase
factor) and an efficient method to obtain it was recently developed \cite{Mozrzymas2014} 
and used in Ref.~\cite{Jocic2020} to transform the effective model into the desired form.
The procedure described in Ref.~\cite{Mozrzymas2014} is exact but relies on a critical step where one has to find for which indices $(a,b)$ the weight matrix $r_{a,b}$ is finite. For transformations between irreps, any of the finite $r_{a,b}$ lead to equivalent unitary transformations. However, for transformations between reducible representations, one needs to identify, within the set of finite $r_{a,b}$, the ones that yield nonequivalent transformation matrices that combine to form the final transformation matrices $U$.
This can be a complicated numerical task.
Here, instead, we propose an alternative method that applies more easily to reducible representations and allows us to obtain the transformation matrix $U$ with a systematic approach. Next, we describe the method, and later in Sec.~\ref{sec:spinfulgraphene} we illustrate its capabilities using the spinful graphene example.

The set of unitary transformations $\mathcal{O}^{i}=U\cdot \mathcal{C}^{i}\cdot U^{\dagger}$
for each $S_{i}\in\mathcal{G}_{\bm{k}_{0}}$ compose a system of equations
for $U$. These can be written in terms of its matrix elements
in a linearized form that reads as 
\begin{equation}
    \label{eq:UABU}
    \sum_{j}U_{m,j}\mathcal{C}_{j,n}^{i}-\mathcal{O}_{m,j}^{i}U_{j,n}=0.
\end{equation}
Defining a vector $\ensuremath{\bm{V}=\{U_{1,1},U_{1,2},\cdots,U_{2,1},\cdots,U_{N,N}\}}^{T}$,
where $N$ is the order of the representations (number of bands in
set $A$), allow us to cast the equation above as $\bm{Q}_{i}\cdot\bm{V}=0$,
with $\bm{Q}_{i}=1_{N}\otimes(\mathcal{C}^{i})^{T}-\mathcal{O}^{i}\otimes1_{N}$ of size
$N^{2}\times N^{2}$, and $1_{N}$ as the $N\times N$ identity matrix.
Since the same similarity transformation $U$ must apply for all $S_{i}$,
we stack each $\bm{Q}_{i}$ into a rectangular matrix
$\bm{Q}=[\bm{Q}_{1},\bm{Q}_{2},\cdots,\bm{Q}_{q}]^T$ of size $(qN^{2})\times N^{2}$. 
The full set of equations now read as $\bm{Q}\cdot\bm{V}=0$, such that
the solution $\bm{V}=\sum_{j=1}^{N_{Q}}c_{j}\bm{v}_{j}$ is a linear
combination of the nullspace $\{\bm{v}_{j}\}$ of $\bm{Q}$, with
coefficients $c_{j}$ and nullity $N_{Q}$. The matrix $U$ can be
recovered from the elements of $\bm{V}$, which follow from its definition
above. If $u_{j}$ is the matrix reconstructed form of $\bm{v}_{j}$,
we can write $U=\sum_{j=1}^{N_{Q}}c_{j} u_{j}$. 

Additionally, it is interesting to consider anti-unitary symmetries.
These can be either the time-reversal symmetry (TRS) itself, or combinations
of TRS and space group operations (magnetic symmetries) \cite{Tinkham2003,Dresselhaus2007}.
For instance, in spinful graphene neither TRS nor spatial inversion
are symmetries of the K point, but their composition is an important
symmetry that enforces a constraint on the allowed SOC terms (see Sec.~\ref{sec:spinfulgraphene}).
Following a notation
similar to the one above, let us refer to these magnetic symmetries
as $\bar{\mathcal{C}}^{i}=D^{\mathcal{C}}(\bar{S}_{i})\mathcal{K}\equiv\tilde{\mathcal{C}}^{i}\mathcal{K}$
and $\bar{\mathcal{O}}^{i}=D^{\mathcal{O}}(\bar{S}_{i})\mathcal{K}\equiv\tilde{\mathcal{O}}^{i}\mathcal{K}$,
where $\mathcal{K}$ is the complex conjugation, and $(\tilde{\mathcal{C}}^{i},\tilde{\mathcal{O}}^{i})$
are the unitary parts of $(\bar{\mathcal{C}}^{i},\bar{\mathcal{O}}^{i})$. Now the basis
transformation for these symmetries read as $\tilde{\mathcal{O}}^{i}=U^{*}\cdot\tilde{\mathcal{C}}^{i}\cdot U^{\dagger}$,
where we choose to apply $\mathcal{K}$ to the left (this choice is
for compatibility with the python package IrRep \cite{IrRep}). To
add this equation to the $\bm{Q}$ matrix above, we consider $U$
and $U^{*}$ as independent variables. Then, as above, it follows
the linearized form
\begin{align}
    \sum_{j}U_{m,j}^{*}\tilde{\mathcal{C}}_{j,n}^{i}-\tilde{\mathcal{O}}_{m,j}^{i}U_{j,n} & =0.
    \label{eq:UABU2}
\end{align}

In all cases, the expression for the transformation matrix is $U=\sum_{j=1}^{N_{Q}}c_{j} u_{j}$,
where the coefficients $c_{j}$ are so far undefined.
To find these coefficients $c_{j}$, we numerically minimize the residues 
$R(\{c_{j}\})=\sum_{i}||\mathcal{O}_{i}-U\cdot \mathcal{C}^{i}\cdot U^{\dagger}||^2$,
and $\tilde{R}(\{c_{j}\})=\sum_{i}||\tilde{\mathcal{O}}_{i}-U^{*}\cdot\tilde{\mathcal{C}}^{i}\cdot U^{\dagger}||^2$.
The global minima of these residues, $R(\{c_{j}\}) = \tilde{R}(\{c_{j}\}) \equiv 0$, yields \emph{a} solution $U(\{c_j\})$, such that small perturbations to the coefficients $c_j \rightarrow c_j + \delta c_j$ lead to quadratic deviations from the minima, \textit{e.g.}, $R \propto |\delta c_j|^2$. This procedure opens a question of whether the solution $U(\{c_j\})$ at the global minima is unique.

Since $U$ represents a transformation between two basis sets
(\textit{e.g.}, $\ket{n_\mathcal{O}} =U \ket{n_\mathcal{C}}$),
it expected to be unique. However, the problem here is formulated such that we
explicitly have the eigenstates $\ket{n_\mathcal{C}}$ that compose the crude DFT basis set $\mathcal{C}$,
while for the optimal symmetry-adapted basis set $\mathcal{O}$ we know only
how we expect the eigenstates $\ket{n_\mathcal{O}}$ to transform under the symmetry operations of the group.
Therefore, instead of solving for $U$ directly from the linear basis transformation $\ket{n_\mathcal{O}} = U \ket{n_\mathcal{C}}$, we rely on the quadratic equations for the transformation between the symmetry operators (\textit{e.g.}, $D^{\mathcal{O}}(S)=U\cdot D^{\mathcal{C}}(S)\cdot U^{\dagger}$), or their linearized forms in Eq.~\eqref{eq:UABU} and Eq.~\eqref{eq:UABU2}. First, consider that $\mathcal{O}$ and $\mathcal{C}$ refer to distinct, but equivalent irreps. As emphasized in \cite{Mozrzymas2014}, it follows from Schur's lemma that the transformation $U$ is unique modulo a phase. Indeed, for the unitary constraints, $\mathcal{O}^{i}=U\cdot \mathcal{C}^{i}\cdot U^{\dagger}$, the solution $U$ is invariant under $U\rightarrow e^{i\theta}U$ for any real $\theta$, while for the anti-unitary constraint, $\tilde{\mathcal{O}}^{i}=U^{*}\cdot\tilde{\mathcal{C}}^{i}\cdot U^{\dagger}$, $U$ is invariant only for $\theta=0$ or $\pi$. Next, without loss of generality, let us consider that $\mathcal{O}$ and $\mathcal{C}$ refer to reducible representations already cast in block-diagonal forms. In this case, the solution $U = U_1 \oplus U_2 \oplus \cdots$ also takes a block-diagonal form, where each block $U_j$ corresponds to a transformation within a single irrep subspace. It follows that each $U_j$ is unique modulo the phases above. The overall global phase of $U$ does not affect the calculation of our matrix elements. However, the arbitrary relative phases between the blocks $U_j$
might lead to ill-defined phases of matrix elements between eigenstates of different irreps if the anti-unitary symmetries are not
informed. In contrast, if anti-unitary symmetries are used, the undefined
phase factor in the matrix elements is just a sign.

\subsection{Matrix elements via DFT\label{subsec:matelDFT}}

As shown above, our approach to obtain a $\bm{k}\cdot\bm{p}$ model
directly from the DFT data relies on two quantities: (i) the band
energies $E_{n}^{0}$ at the $\bm{k}\cdot\bm{p}$ expansion point
$\bm{k}_{0}$; and (ii) the matrix elements $\bm{P}_{m,n}=\bra{m}\bm{\pi}\ket{n}$
also calculated at $\bm{k}_{0}$ for all bands $\{\ket{n}\}$. The
band energies $E_{n}^{0}$ are a straightforward output of any DFT
code. Therefore, here we discuss only the calculation of $\bm{P}_{m,n}=\bra{m}\bm{\pi}\ket{n}$.

We focus on the Quantum ESPRESSO (QE) \cite{Giannozzi2009,Giannozzi2017} implementation of \emph{ab initio} DFT \cite{dft1,dft2}. There, the Hamiltonian is split into the core and intercore regions via the Projector Augmented Wave (PAW) method \cite{Bloechl1994,Kresse1999,Corso2010}, which is backward compatible with ultrasoft (USPPs) \cite{Vanderbilt1990,Kresse1999} and norm-conserving pseudo-potentials (NCPP) \cite{Hamann1979,Bachelet1982,Hamann2013}. In these approaches, the atomic core region is replaced by pseudopotentials, which are constructed from single-atom DFT simulations with the Dirac equation in the scalar relativistic or full relativistic approaches. Thus, for molecules or crystals, QE solves a pseudo-Schr\"odinger equation, with the atomic potentials replaced by the pseudopotentials. Here we shall not go through the details of the PAW and pseudopotential methods. For the interested reader, we suggest Refs. \cite{Bloechl1994,Kresse1999,Corso2010}. Instead, for now, it is sufficient to conceptually understand that QE provides numerical solutions for the Schr\"odinger equation with the fine structure corrections, which can be expressed by the Hamiltonian
\begin{equation}
    H\approx p^{2}+V(\bm{r})+H_{{\rm SR}}+\frac{\alpha^{2}}{4}(\bm{\sigma}\times\nabla V)\cdot\bm{p},
\end{equation}
where $H_{{\rm SR}}=H_{{\rm D}}+H_{{\rm MV}}$ contain the Darwin
and mass-velocity contributions, as presented above, and the last
term is the spin-orbit coupling.

\subsubsection{Matrix elements of the velocity}

Fortunately, the QE code already provides tools to calculate the matrix
elements of the velocity operator $\frac{1}{2}\bm{v}=\frac{i}{2}[H,\bm{r}]$,
which reads as
\begin{align}
    \label{eq:velocity}
    \frac{\bm{v}}{2} & =\frac{1}{2}\frac{\partial H}{\partial\bm{p}}=\bm{\pi}+\frac{1}{2}\frac{\partial H_{{\rm MV}}}{\partial\bm{p}}\approx\bm{\pi},
\end{align}
where we neglect the mass velocity corrections (see Appendix \ref{sec:massvelocity}).
Thus, we find that $\bm{P}_{m,n}=\bra{m}\bm{\pi}\ket{n}\approx\bra{m}\frac{1}{2}\bm{v}\ket{n}$.
The calculation of $\bm{P}_{m,n}$ is already partially included in
the post-processing tool \texttt{bands.x} (file \texttt{PP/src/bands.f90}),
within the \texttt{write\_p\_avg} subroutine (file \texttt{PP/src/write\_p\_avg.f90}).
This calculation includes the necessary PAW, USPPs, or NCPPs corrections,
which are critical for materials where the wavefunction strongly
oscillates near the atomic cores \cite{Kageshima1997}. However,
the \texttt{write\_p\_avg} subroutine only calculates $|\bm{P}_{m,n}|^{2}$
for $m$ in the valence bands (below the Fermi level) and $n$ in
the conduction bands (above the Fermi level). To overcome this limitation,
we have built a patch that modifies \texttt{bands.f90} and \texttt{write\_p\_avg.f90}
to calculate $\bm{P}_{m,n}$ for all bands. This leads to a modified
\texttt{bands.x} with options to follow with its original behavior
or to calculate $\bm{P}_{m,n}$ according to our needs. This is
controlled by a new flag \texttt{lpall = False/True}
added to the input file of \texttt{bands.x} in addition
to the \texttt{lp = True}.
Its default value (\texttt{lpall = False}) runs \texttt{bands.x} with
its original code, while the option \texttt{lpall = True} 
instructs \texttt{bands.x} to store all $\bm{P}_{m,n}$
into the file indicated by the input parameter \texttt{filp}.

In general, it is preferable to patch QE to use the full $\bm{P}_{m,n}$, since the calculation is faster and more precise.
Nevertheless, if the user prefers not to apply our patch to modify
QE, our code can calculate an approximate $\bm{P}_{m,n}$ using only
the plane-wave components outputted by the QE code. In this case,
we consider that the pseudo-wavefunction is a reasonable approximation
for the all-electron wavefunction, thus neglecting PAW corrections,
which are necessary to account for SOC. Therefore,
under this approximation, $\bm{P}_{m,n}\approx\bra{m}\bm{p}\ket{n}$.
The relevance of these PAW/SOC corrections to $\bm{P}_{m,n}$ are
presented in the example shown in Sec.~\ref{sec:PSOC}.
Within this approximation, the wavefunction
$\psi_{n,\bm{k}}(\bm{r})$ for the band $n$ at quasi-momentum $\bm{k}$,
and $\bm{P}_{m,n}$ read as
\begin{align}
    \psi_{n,\bm{k}}(\bm{r}) & \approx\frac{1}{\sqrt{\Omega}}\sum_{\bm{G}}c_{n}(\bm{G})e^{i(\bm{k}+\bm{G})\cdot\bm{r}},
    \label{eq:psiQE}
    \\
    P_{m,n} & \approx\sum_{\bm{G}}(\bm{k}+\bm{G})c_{m}^{\dagger}(\bm{G})c_{n}(\bm{G}),
    \label{eq:PmnNoPAW}
\end{align}
where $c_{n}(\bm{G})$ are the plane-wave expansion coefficients (spinors
in the spinful case),  $\Omega$ is the normalization volume, and $\bm{G}$
are the lattice vectors in reciprocal space. To implement this calculation,
and the one shown next, we use the \texttt{IrRep} python package \cite{IrRep},
since it already has efficient routines to read and manipulate the
QE data. 

\subsubsection{Matrix elements of the symmetry operators}
\label{sec:Pmn}

To calculate the matrix elements of the symmetry operators, it is
sufficient to consider $\psi_{n,\bm{k}}(\bm{r})$ from Eq.~\eqref{eq:psiQE}.
In this case, it is safe to neglect PAW corrections, since they must
transform identically to the plane-wave parts under the symmetry operations
of the crystal space group. For a generic symmetry operation $S\in\mathcal{G}_{\bm{k}_{0}}$,
its matrix elements read as
\begin{multline}
    D_{m,n}^{\psi}(S)=\sum_{\bm{G},\bm{G}'}c_{m}^{\dagger}(\bm{G}')c_{n}(\bm{G})
    \\
    \int e^{-i(\bm{k}+\bm{G}')\cdot\bm{r}}e^{-iS^{-1}(\bm{k}+\bm{G}')\cdot\bm{r}} \dfrac{d^{3}r}{\Omega}.
\end{multline}
Using the plane-wave orthogonality, one gets
\begin{equation}
    D_{m,n}^{\psi}(S)=\sum_{\bm{G}}c_{m}^{\dagger}\big(-\bm{k}+S^{-1}\cdot(\bm{k}+\bm{G})\big)c_{n}(\bm{G}),
    \label{eq:Smn}
\end{equation}
where $S^{-1}$ is the inverse of $S$, and $S^{-1}\cdot(\bm{k}+\bm{G})$
is its action on the $(\bm{k}+\bm{G})$ vector. For instance, if $S=I$
is the spatial inversion symmetry, $S^{-1}\cdot(\bm{k}+\bm{G})=-\bm{k}-\bm{G}$,
and $D_{m,n}^{\psi}(S)=\sum_{\bm{G}}c_{m}^{\dagger}(-2\bm{k}-\bm{G})c_{n}(\bm{G})$. 

\section{Hands-on example: graphene}
\label{sec:graphene}

In this section, we present a detailed example and results for spinless graphene, and a shorter discussion on spinful graphene in Sec.~\ref{sec:spinfulgraphene} to illustrate the case of transformations between reducible representations.
Graphene \cite{Novoselov2004, Novoselov2005} is nowadays one of the most studied materials due to the discovery of its Dirac-like effective low energy model, which reads as $H=\hbar v_{F}\bm{\sigma}\cdot\bm{k}$. 
Here, the $\bm{\sigma}$ Pauli matrices act on the orbital pseudo-spin subspace,
$\bm{k}=(k_{x},k_{y})$ is the quasi-momentum, and $v_{F}$ is the Fermi velocity, which is the unknown coefficient that we want to calculate in this example. For this purpose, we follow a pedagogical route in this first example. First, we present the symmetry characteristics of the graphene lattice and its wavefunctions at the K point. Then, we show the results for the representation matrices and Hamiltonian in the crude and optimal symmetry-adapted basis to illustrate how the symmetry optimization of Section \ref{subsec:symmopt} is used to build the optimal symmetry-adapted Hamiltonians and identify the numerical values for its coefficients. Later, in Section \ref{subsec:Running-the-code} we show a step-by-step tutorial on how to run the code. This example was chosen for its simplicity, which allows for a clear discussion of each step. Later, in Section \ref{sec:Examples} we present a summary of examples for other materials of current interest.

\begin{figure}[th]
    \includegraphics[width=1\columnwidth]{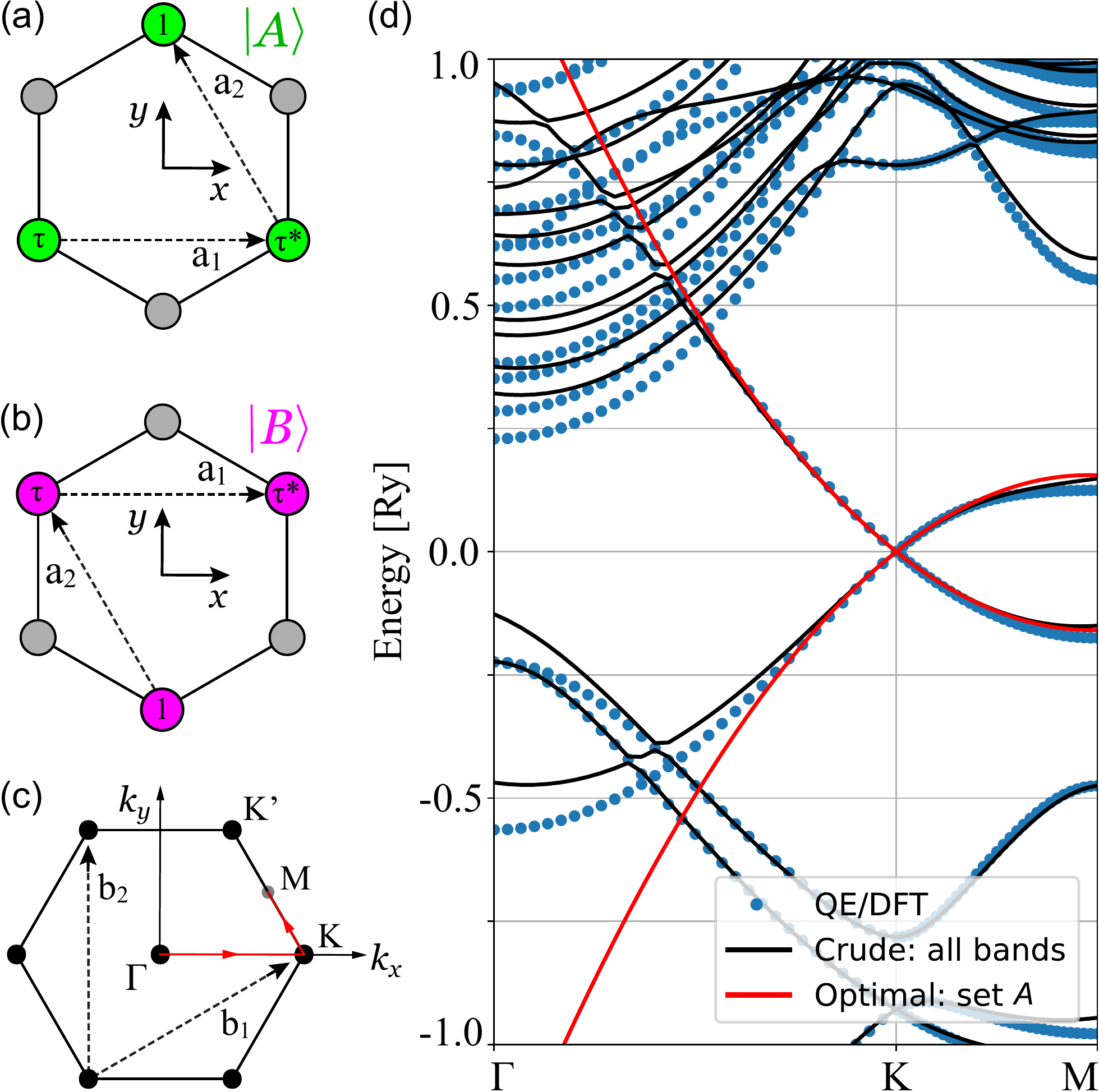}
    \caption{\label{fig:graphene}Graphene lattices emphasizing the Dirac cone
    eigenstates at the K point, where (a) $\ket{A}=\ket{(X+iY)Z}$ and
    (b) $\ket{B}=\ket{(X-iY)Z}$. Both eigenstates are composed by $p_{z}$
    orbitals centered at the colored sites (A and B lattices) with the
    Bloch phase factors indicated within the circles, where $\tau=\exp(i2\pi/3)$.
    (c) The first Brillouin zone, marking the path $\Gamma-K-M$ used to
    plot the bands in (d). (d) Band structure for graphene calculated
    via QE/DFT (blue circles), all bands model [Eq.~\eqref{eq:crude_all_bands}]
    (black lines), and optimal symmetry-adapted model [Eq.~\eqref{eq:Hgraphene}] for
    the two bands forming the Dirac cone (red). Here, the QE/DFT simulation
    was performed with 300 bands.}
\end{figure}

Before discussing the details, we summarize the results for the band
structure of graphene in Fig.~\ref{fig:graphene}, which compares
the DFT data with our two main models. The black lines are calculated
from the all bands model from Eq.~\eqref{eq:crude_all_bands},
which uses the matrix elements
$\bm{P}_{m,n}$ in the original crude DFT basis without
further processing. In contrast, the red lines are the band structure
calculated with the folded-down Hamiltonian for a set $A$ composed
by the two bands near the Fermi energy that defines the Dirac cone,
and considers the symmetry optimization process to properly identify
the $\bm{k}\cdot\bm{p}$ parameters. This optimal symmetry-adapted Hamiltonian is shown
in Eq.~\eqref{eq:Hgraphene} below, and the numerical value for its
parameters is shown at \emph{Step 7} in Section \ref{subsec:Running-the-code}. 

\subsection{Overview of the theory and symmetry optimization}

The crystal structure of graphene is a hexagonal monolayer of carbon
atoms, as shown in Figs. \ref{fig:graphene}(a) and \ref{fig:graphene}(b),
which is invariant under the P6/mmm space group (\#191). However,
since its Dirac cone is composed of $p_{z}$ orbitals only, it is sufficient
to consider the $C_{6V}$ factor group to describe the lattice. Particularly, at the K point [see Fig.~\ref{fig:graphene}(c)],
the star of K corresponds to the little group $C_{3V}$, which is
generated by a 3-fold rotation $C_{3}(z)$ and a mirror $M_{y}$.
The Dirac bands of graphene are characterized by the irrep $E$ of
$C_{3V}$ (or irrep $K_{6}$ from P6/mmm \cite{Elcoro2017}), which
is composed by basis functions $(xz,yz)$. 

To build the optimal symmetry-adapted effective model via the method of invariants,
we need to specify a basis and calculate the matrix representation
of the symmetry operations mentioned above. Since the wavefunctions
of the Dirac cone transform as the irrep $E$ of $C_{3V}$, a naive
choice would be $A_{{\rm unconv}}=$$\{\ket{XZ},$$\ket{YZ}\}$,
which corresponds to a set $A$ in Section \ref{subsec:kp}.
This choice of basis refers to a possible $\mathcal{C}$ representation in Section
\ref{subsec:symmopt}, and it yields
\begin{align}
    D^{{\rm unconv}}(C_{3}(z)) & =
    \begin{pmatrix}
        \cos\theta & -\sin\theta\\
        \sin\theta & \cos\theta
    \end{pmatrix},
    \\
    D^{{\rm unconv}}(M_{y}) & =
    \begin{pmatrix}
        1 & 0 \\
        0 & -1
    \end{pmatrix},
    \\
    H_{{\rm unconv}} & \approx
    \begin{pmatrix}
        c_{0}-c_{1}k_{x} & c_{1}k_{y} \\
        c_{1}k_{y} & c_{0}+c_{1}k_{x}
    \end{pmatrix},
\end{align}
where $\theta=2\pi/3$. Here $H_{{\rm unconv}}$ is obtained via \texttt{Qsymm}
up to linear order in $\bm{k}$, for brevity. While the eigenenergies
of $H_{{\rm unconv}}$ represent correctly the Dirac cone as $E_{\pm}=c_{0}+|c_{1}|\sqrt{k_{x}^{2}+k_{y}^{2}}$,
the Hamiltonian $H_{{\rm unconv}}$ takes an undesirable unconventional
form. 

A more convenient choice is $A_{{\rm conv}}=\{\ket{(X+iY)Z},$$\ket{(X-iY)Z}\}$,
which is illustrated in Figs. \ref{fig:graphene}(a) and \ref{fig:graphene}(b).
This choice of basis leads to
\begin{align}
    D^{{\rm conv}}(C_{3}(z)) & =
    \begin{pmatrix}
        e^{i\theta} & 0\\
        0 & e^{-i\theta}
    \end{pmatrix},
    \\
    D^{{\rm conv}}(M_{y}) & =
    \begin{pmatrix}
        0 & 1\\
        1 & 0
    \end{pmatrix},
    \\
    H_{{\rm conv}} & \approx
    \begin{pmatrix}
        c_{0} & c_{1}k_{-}\\
        c_{1}k_{+} & c_{0}
    \end{pmatrix}+
    \begin{pmatrix}
        c_{2}k^{2} & c_{3}k_{+}^{2}\\
        c_{3}k_{-}^{2} & c_{2}k^{2}
    \end{pmatrix},
    \label{eq:Hgraphene}
\end{align}
where $k_{\pm}=k_{x}\pm ik_{y}$. Now, up to linear order in k, we
see that $H_{{\rm conv}}\approx c_{0}+c_{1}\bm{\sigma}\cdot\bm{k}$,
where $\bm{\sigma}$ act on the subspace set by $A_{{\rm conv}}$,
and we identify $c_{1}=\hbar v_{F}$. Additionally, the k-quadratic
terms that lead to trigonal warping corrections. Notice that both
choices, $A_{{\rm unconv}}$ and $A_{{\rm conv}}$, are
equivalent representations, but the conventional one leads to the
familiar \emph{form} of the graphene Hamiltonian. These two basis
sets are related by an unitary transformation $U$, such that $A_{{\rm conv}}=U \cdot A_{{\rm unconv}}$
and $H_{{\rm conv}}=UH_{{\rm unconv}}U^{\dagger}$, with
\begin{equation}
    U_{{\rm unconv}\rightarrow{\rm conv}}=\frac{1}{\sqrt{2}}
    \begin{pmatrix}
        1 & i\\
        1 & -i
    \end{pmatrix}.
\end{equation}

Next, let us analyze the set $A_{{\rm QE}}$ of numerical wavefunctions
from QE. Do they correspond to $A_{{\rm QE}} = A_{{\rm conv}}$
or $A_{{\rm QE}} = A_{{\rm unconv}}$? The answer is neither.
Since it is a raw numerical calculation, typically diagonalized via
the Davidson algorithm \cite{Davidson1975}, a degenerate or nearly
degenerate set of eigenstates might be in any linear combination of
its representative basis. Therefore, the symmetry optimization step
is essential to find the matrix transformation $U$ that yields $A_{{\rm conv}}=U\cdot A_{{\rm QE}}$.
To visualize this, let us check the matrix representations of the
symmetry operators above, and the effective Hamiltonian calculated
from the crude QE data. For the symmetry operators, we find
\begin{align}
    D^{{\rm QE}}(C_{3}(z)) & \approx
    \begin{pmatrix}
        -0.5 & -0.35+0.79i\\
        0.35+0.79i & -0.5
    \end{pmatrix},
    \\
    D^{{\rm QE}}(M_{y}) & \approx
    \begin{pmatrix}
        +0.5 & 0.35-0.79i\\
        0.35+0.79i & -0.5
    \end{pmatrix},
\end{align}
While this cumbersome numerical representation does not resemble neither
$A_{{\rm conv}}$ nor $A_{{\rm unconv}}$, our symmetry
optimization process correctly finds a transformation matrix $U$
that returns $A_{{\rm conv}}=U\cdot A_{{\rm QE}}$, where
\begin{equation}
    U\approx
    \begin{pmatrix}
        0.7i & -0.28+0.65i\\
        -0.6+0.37i & 0.7-0.1i
    \end{pmatrix}.
\end{equation}
Finally, for the Hamiltonian, up to linear order in k and in the original
QE basis, we find
\begin{multline}
    H_{{\rm QE}}\approx
    \begin{pmatrix}
        -0.37 & -0.25+0.57i\\
        -0.25-0.57i & 0.37
    \end{pmatrix} k_{x}
    \\
    +\begin{pmatrix}
        0.62 & 0.15-0.34i\\
        0.15+0.34i & 0.62
    \end{pmatrix} k_{y},
\end{multline}
which takes a cumbersome \textit{form} in this raw numerical basis. However, applying the transformation $U$, the symmetry adapted model
becomes
\begin{align}
    H_{N\times N}^{{\rm optimal}}=UH_{N\times N}^{{\rm DFT}}U^{\dagger}\approx0.72\,\bm{\sigma}\cdot\bm{k}.
\end{align}
Here we identify $\hbar v_{F}=0.72$ in Rydberg units, yielding $v_{F}=0.83\times10^{6}$
m/s. The resulting band structure calculated from $H^{{\rm optimal}}$,
including the k-quadratic terms, is shown as red lines in Fig.~\ref{fig:graphene}(d)
and it matches well the QE/DFT data near K.

\subsection{Running the code\label{subsec:Running-the-code}}

The example presented here is available in the \texttt{Examples/graphene-nosoc.ipynb}
notebook in the code repository, and shown in Algorithm \ref{alg:minimal}.
Here we show only the minimal procedure to read the DFT data, build
an effective model from the symmetry constraints, and calculate the
numerical values for the model parameters. Complementary, the full
code in \texttt{Examples/graphene-nosoc.ipynb} shows how to plot the
data presented in our figures.

For now, we assume that the DFT simulation was successful. The suggested
steps to run QE and prepare the data for our code is to run the \texttt{calculation=`scf'}
and \texttt{calculation=`bands'} with \texttt{pw.x}. Then, run
\texttt{bands.x} to extract the bands from QE's output
and store it in \texttt{gnuplot} format to plot the figures. Here,
for graphene, we assume that the \texttt{bands} calculation was run
for a path $\Gamma-{\rm K}-{\rm M}$ with 30 points between each section,
such that K is the 31st point in the list. \newline

\noindent Next, we describe each step shown in Algorithm \ref{alg:minimal}. \newline

\paragraph*{Step 1.}

After running QE, the first step is to read the DFT data
from the QE's output folder. The command \texttt{dft2kp.irrep(...)
}uses the python package \texttt{IrRep} \cite{IrRep} to read the
data for the selected k point to be used in the $\bm{k}\cdot\bm{p}$
expansion, as indicated by the parameters \texttt{kpt} and \texttt{kname}.
The data is read from the folder indicated by the parameter \texttt{dftdir},
while \texttt{outdir} and \texttt{prefix} refer to values used in
the input file of QE's \texttt{pw.x} calculation. Additionally, the
command \texttt{dft2kp.irrep(...)} also accepts extra parameters from
the package \texttt{IrRep} (see code documentation). \newline

\paragraph*{Step 2.}

In step 2, the code will either read or calculate the matrix elements
$\bm{P}_{m,n}$ to build the effective models. If the user runs QE
modified by our patch, the QE tool \texttt{bands.x} will generate
a file \texttt{kp.dat} that already contains the values for $\bm{P}_{m,n}$.
In this case, the user must inform the name of this file via the parameter
\texttt{qekp}. Otherwise, if \texttt{qekp} is omitted, our code calculates
an approximate value for $\bm{P}_{m,n}\approx\bra{m}\bm{p}\ket{n}$
from the pseudo-wavefunction of QE, as in Eq.~\eqref{eq:PmnNoPAW},
which neglects all SOC corrections.  \newline

\paragraph*{Step 3.}

Next, the user must choose which set of bands will be considered to
build the model. This is the set $A$ in Section \ref{subsec:kp}.
In this example, we select bands 3 and 4, which correspond to the Dirac
cone of graphene. The code analyzes the list of bands and identifies
their irreducible representations (irreps) using the \texttt{IrRep}
package \cite{IrRep}. Here, the set $A$ must contain only
complete sets of irreps, otherwise the L\"owdin perturbation theory
would fail with divergences [see Eq.~\eqref{eq:lowdin}], since
the remote bands of set $B$ would have at least one band degenerated
with a band from set $A$. If this condition fails, the code
stops with an error message. Otherwise, if set $A$ is valid,
the code outputs a report indicating the space group of the crystal
(e.g., P6/mmm), the selected set of bands (e.g., [3,4]), their
irrep (e.g., $K_{6}$ \cite{Elcoro2017}), and degeneracy (2). The
report reads as
\begin{minted}[fontsize={\small},breaklines=true,tabsize=2]{tex}
Space group  191: P6/mmm
Verifying set A: [3 4]
Band indices: [3, 4] Irreps: (K6) Degeneracy: 2
\end{minted}
Additionally, in this step, the code also calculates the crude
effective model for the bands in set $A$ via L\"owdin partitioning \cite{Lowdin1951}.
It stores the folded Hamiltonian in a Python dictionary (\texttt{kp.Hdict})
representing the matrices $h_{i,j,l}$ in the crude DFT basis
that define $H^{{\rm DFT}}(\bm{k})=\sum_{i,j,l}h_{i,j,k}k_{x}^{i}k_{y}^{j}k_{z}^{l}$.
For instance, \texttt{kp.Hdict[`xx']} refers to the matrix $h_{2,0,0}$
that defines the term $h_{2,0,0}k_{x}^{2}$. \newline

\paragraph*{Step 4.}

In step 4 we build the optimal symmetry-adapted model using \texttt{Qsymm} \cite{Varjas2018},
which solves Eq.~\eqref{eq:invariants} for the method of invariants.
In Algorithm \ref{alg:minimal}, we build the representations for
the symmetry operations $C_{3}(z)$, $M_{y}$, $M_{z}$, and $\mathcal{TI}$.
Above we have discussed only the first two for simplicity. Here we
also include the mirror $M_{z}$, and the anti-unitary symmetry
$\mathcal{TI}$, which is composed of the product of time-reversal and spatial inversion
symmetries. The mirror $M_{z}$ has a trivial representation $D^{\psi}(M_{z})=-1$,
since the orbitals that compose the Dirac bands in graphene are all
of Z-like (odd in z). The $\mathcal{TI}$ representation follows
from $A_{{\rm conv}}$ presented above by recalling that spinles
time-reversal is simply the complex conjugation and the spatial inversion
takes $(X,Y,Z)\rightarrow(-X,-Y,-Z)$. In this particular example,
the $\mathcal{TI}$ symmetry does not play an important role, but
it is essential for a spinful graphene example, as it constrains the
SOC terms at finite $\bm{k}$ (see Sec.~\ref{sec:spinfulgraphene}).
The command \texttt{dft2kp.qsymm(...)} calls
\texttt{Qsymm} to build the effective model from the list of symmetries,
indicated by \texttt{symm}, up to order $k^{2}$, as indicated by
\texttt{total\_power}. We recommend always using \texttt{dim=3} [three
dimensions for $\bm{k}=(k_{x},k_{y},k_{z})$] because QE always
work with the 3D space groups. Additionally, the command \texttt{dft2kp.qsymm(...)}
accepts other parameters that are given to the \texttt{Qsymm} package
(see code documentation). By default, this command outputs the optimal symmetry-adapted
Hamiltonian, which matches the one in Eq.~\eqref{eq:Hgraphene}. \newline

\paragraph*{Step 5.}

Next, we start the symmetry optimization process. The first call \texttt{kp.get\_symm\_matrices()}
calculates, via Eq.~\eqref{eq:Smn}, the matrix representation for
all symmetry operators identified in the QE data by the \texttt{IrRep}
package. However, neither QE nor \texttt{IrRep} account for the anti-unitary
symmetries. Therefore, we call here the optional routine \texttt{kp.add\_antiunitary\_symm(...)},
which manually adds the anti-unitary symmetry to the list of QE symmetries
and matches it with the corresponding symmetry of \texttt{Qsymm} informed
on its first parameter. In this example, we add the $\mathcal{TI}$
symmetry built with \texttt{Qsymm} above. This operator needs to be
complemented with a possible non-symmorphic translation vector, which
is zero in this case, as shown by the second parameter of \texttt{kp.add\_antiunitary\_symm(...).}
Both calls, \texttt{kp.get\_symm\_matrices()} and \texttt{kp.add\_antiunitary\_symm(...)},
calculate the matrix representations in the crude QE basis. \newline

\paragraph*{Step 6.}

To calculate the transformation matrix $U$, we compare the ideal
matrix representations informed via \texttt{Qsymm} (object \texttt{qs})
and the crude QE matrix representations (object \texttt{kp}).
The call \texttt{dft2kp.basis\_transform(...)} performs this comparison
and returns an error if the symmetries in both objects do not match.
More importantly, it calculates the transformation matrix $U$ solving
Eq.~\eqref{eq:UABU} and Eq.~\eqref{eq:UABU2}. The matrix $U$ is
stored in the object \texttt{optimal.U}. If the calculation of $U$
is successful, the code applies $U$ to rotate the $h_{i,j,l}$ terms
in \texttt{kp.Hdict} from the crude DFT basis into the optimal symmetry-adapted
basis. This allows for direct identification of the coefficients
$c_{n}$ from Eq.~\eqref{eq:Hgraphene}, which are stored in \texttt{optimal.coeffs}.
Additionally, the code builds the numerical optimal symmetry-adapted model and provides
a callable object \texttt{optimal.Heff(kx, ky, kz)} that returns the
numerical Hamiltonian $H_{N\times N}^{{\rm optimal}}$ for
a given value of $\bm{k}=(k_{x},k_{y},k_{z})$. \newline

\paragraph*{Step 7.}

At last, the code prints a report with the numerical values for the
coefficients $c_{n}$, which are summarized in Table \ref{tab:graphene-nosoc}.
As mentioned above, here we identify $\hbar v_{F}=0.72\text{ a.u.}$,
yielding $v_{F}=0.83\times10^{6}$ m/s after converting the units.

\newpage

\begin{algorithm}
\begin{minted}[numbers=left,fontsize={\small},breaklines=true,tabsize=2]{python}
import numpy as np
import pydft2kp as dft2kp

# import s0, sx, sy, sz: Pauli matrices
from pydft2kp.constants import s0, sx, sy, sz

# step 1: read DFT data
kp = dft2kp.irrep(dftdir='graphene-nosoc',
                  outdir='outdir',
                  prefix='graphene',
                  kpt=31,           
                  kname='K')

# step 2: read or calculate matrix elements of p
kp.get_p_matrices(qekp='kp.dat')

# step 3: define the set alpha
#         applies fold down via Löwdin
setA = [3, 4]
kp.define_set_A(setA)

# step 4: builds optimal model with qsymm
phi = 2*np.pi/3
U = np.diag([np.exp(1j*phi), np.exp(-1j*phi)])
C3 = dft2kp.rotation(1/3, [0,0,1], U=U)
My = dft2kp.mirror([0,1,0], U=sx)
Mz = dft2kp.mirror([0,0,1], U=-s0)
TI = dft2kp.PointGroupElement(R=-np.eye(3), 
                              conjugate=True, 
                              U=sx)
symms = [C3, My, Mz, TI]
qs = dft2kp.qsymm(symms, total_power=2, dim=3);

# step 5: calculate the representation matrices
kp.get_symm_matrices()
# (optional): adds anti-unitary symmetry
kp.add_antiunitary_symm(TI, np.array([0,0,0]))

# step 6: calculates and applies 
#         the transformation U
optimal = dft2kp.basis_transform(qs, kp)

# step 7: print results
optimal.print_report(sigdigits=3)
\end{minted}
\caption{\label{alg:minimal}Minimal example for spinless graphene.}
\end{algorithm}

\begin{table}[H]
    \renewcommand*{\arraystretch}{1.5}
    \caption{\label{tab:graphene-nosoc}Graphene parameters for the Hamiltonian of Eq.~\eqref{eq:Hgraphene}.}
    \begin{ruledtabular}
    \begin{tabular}{c|c|c}
    Coefficient & Values in a.u. & Values in (eV, nm)\\
    \hline
    $c_{0}$ & $\sim0$ & $\sim0$ ${\rm eV}$\\
    $c_{1}$ & $0.72$ & $0.52$ ${\rm eV\,nm}$\\
    $c_{2}$ & $\sim0$ & $\sim0$ ${\rm eV\,nm}^{2}$\\
    $c_{3}$ & $0.82$ & $0.031$ ${\rm eV\,nm}^{2}$\\
    \end{tabular}
    \end{ruledtabular}
\end{table}

\subsection{Spinful graphene}
\label{sec:spinfulgraphene}

To complement the example above, we consider now the spinful graphene (full code available at \texttt{Examples/graphene.ipynb} \cite{DFT2KP}). In this case, due to the small spin-orbit coupling of graphene, the numerical DFT basis functions from QE mix two nearly degenerate irreps into an unintended reducible representation. Nevertheless, our symmetry optimization procedure can properly block diagonalize the symmetry operators according to the intended representation.

\begin{figure}[th]
    \includegraphics[width=\columnwidth]{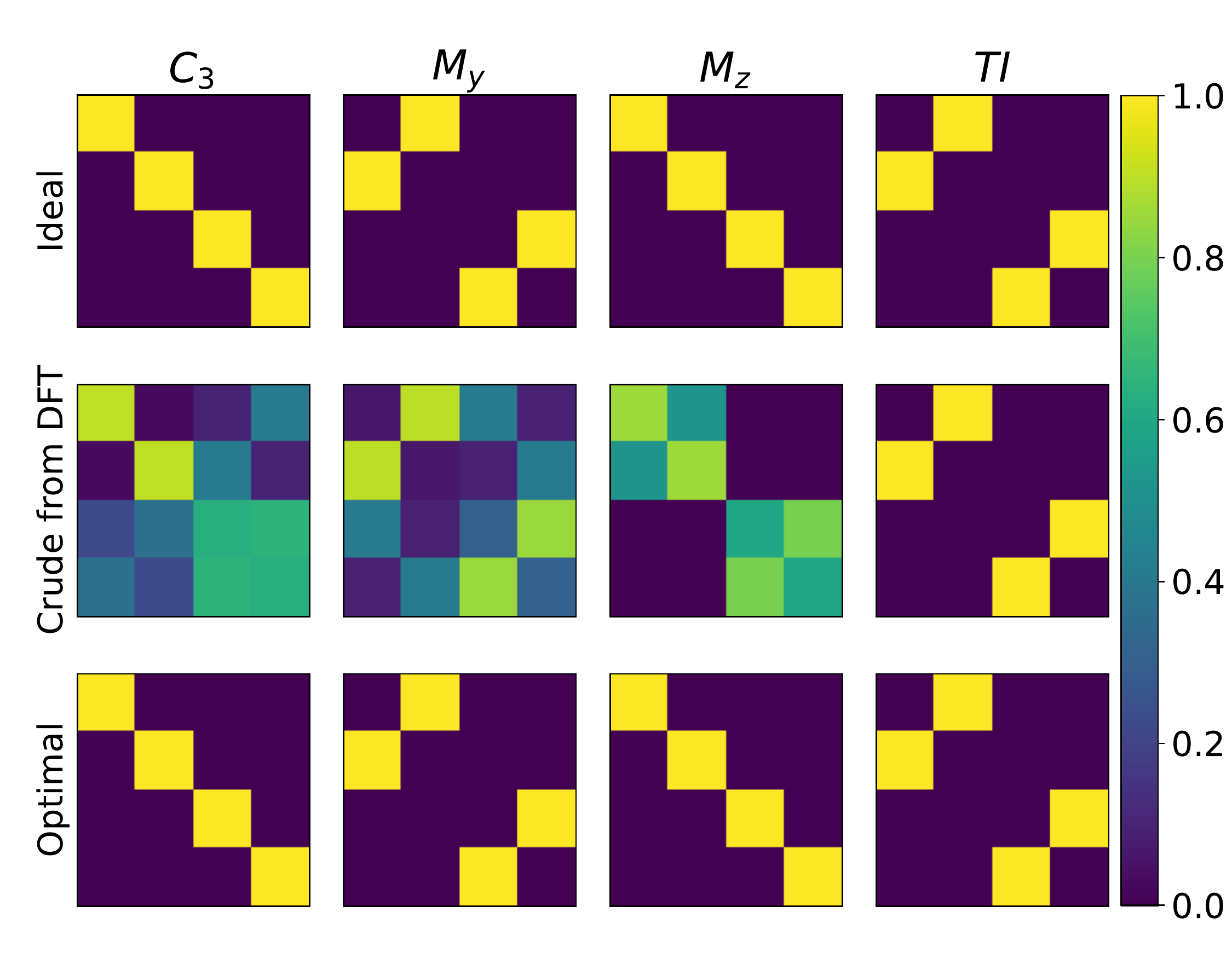}
    \caption{\label{fig:spinful}
    The absolute value of the representation matrices of the symmetry operations for the spinful graphene example, as labeled on top of each column.
    The top line of matrices are defined under the ideal basis informed by the user, i.e. $\{\ket{(X+iY)Z,\uparrow}$, $\ket{(X-iY)Z,\downarrow}$, $\ket{(X-iY)Z,\uparrow}$, $\ket{(X+iY)Z,\downarrow}\}$, as discussed in the text.
    The central line shows the calculated representation matrices under the crude DFT basis from QE, which does not split into the ideal block-diagonal form due to the small SOC gap between the bands.
    Applying our transformation $U$ to the crude representation from the central line, we obtain the optimal symmetry-adapted basis that lead to the proper block-diagonal form of the representation matrices shown in the bottom line.
    }
\end{figure}

To see this, let us first establish the ideal basis in proper ordering that leads to the block-diagonal form of the symmetry operators $C_3(z)$, $M_y$, $M_z$, and $\mathcal{TI}$ (considering the group generators only). Thus, considering the spin, the basis functions now read as $\{\ket{(X+iY)Z,\uparrow}$, $\ket{(X-iY)Z,\downarrow}$, $\ket{(X-iY)Z,\uparrow}$, $\ket{(X+iY)Z,\downarrow}\}$. Under the P6/mmm double space group \cite{Koster, Elcoro2017}, this set of basis functions transform as the sum of two bidimensional irreps \cite{C3vDouble}, namely $\bar{K}_7 \oplus \bar{K}_9$. Under this basis, the symmetry operators listed above take a block-diagonal form, which are illustrated in the top row of Fig.~\ref{fig:spinful}. Algebraically, these read

\newpage

\begin{align}
    D^{\rm ideal}(C_3) &= 
    \begin{pmatrix}
        -\tau^* & 0 & 0 & 0 \\
        0 & -\tau & 0 & 0 \\
        0 & 0 & -1 & 0 \\
        0 & 0 & 0 & 1
    \end{pmatrix},
    \\
    D^{\rm ideal}(M_y) &= 
    \begin{pmatrix}
        0 &-1 & 0 & 0 \\
        1 & 0 & 0 & 0 \\
        0 & 0 & 0 &-1 \\
        0 & 0 & 1 & 0
    \end{pmatrix},
    \\
    D^{\rm ideal}(M_z) &= 
    \begin{pmatrix}
        i & 0 & 0 & 0 \\
        0 &-i & 0 & 0 \\
        0 & 0 & i & 0 \\
        0 & 0 & 0 &-i
    \end{pmatrix},
    \\
    D^{\rm ideal}(\mathcal{TI}) &= 
    \begin{pmatrix}
        0 & 1 & 0 & 0 \\
       -1 & 0 & 0 & 0 \\
        0 & 0 & 0 & 1 \\
        0 & 0 &-1 & 0
    \end{pmatrix}K.
\end{align}

In contrast to the block diagonal form of the $D^{\rm ideal}(\cdots)$ matrices above, the representation matrix for the $C_3(z)$ calculated with the crude DFT basis from QE takes the form
\begin{multline}
    D^{\rm QE}(C_3) \approx 
    \\
    \begin{pmatrix}
        -0.9-0.1i & -0.0-0.0i & +0.1+0.1i & -0.3-0.3i \\
        +0.0-0.0i & -0.9+0.1i & -0.3+0.3i & -0.1+0.1i \\
        -0.2-0.1i & +0.0-0.4i & +0.4+0.5i & +0.3+0.6i \\
        +0.0+0.4i & +0.2-0.1i & -0.3+0.6i & +0.4-0.5i
    \end{pmatrix}.
\end{multline}
Similarly, the crude DFT representation for $M_y$, $M_z$ and $\mathcal{TI}$ also show non-block-diagonal forms in the central line of Fig.~\ref{fig:spinful}. 

The algorithm described in Sec.~\ref{subsec:symmopt} builds a system of equations to find the transformation matrix $U$ that yields $D^{\rm ideal}(S) = U D^{\rm QE}(S) U^\dagger$ for all symmetry $S$ of the group (i.e., $S = \{C_3(z), M_y, M_z, \mathcal{TI}\}$ in this example). The Python code to implement this procedure is nearly identical to Algorithm \ref{alg:minimal}, requiring only (i) the expansion of \texttt{setA}, in \emph{Step 3}, to account for the 4 bands that compose the spinful Dirac cone (i.e., \texttt{setA = [6, 7, 8, 9]} in this Example); and (ii) the replacement of the symmetry matrices from \emph{Step 4} for the ones listed above. From these, in \emph{Step 6} we find the transformation matrix
\begin{multline}
    U \approx 
    \\
    \begin{pmatrix}\small
        +0.1-0.0i & -0.1+0.2i & -0.6-0.6i & -0.4-0.2i \\
        +0.1-0.2i & -0.0-0.1i & +0.4+0.2i & -0.9-0.0i \\
        +0.2+0.2i & -0.9+0.1i & -0.1+0.2i & +0.0+0.1i \\
        -0.6-0.7i & -0.3+0.0i & -0.1+0.1i & +0.1-0.2i
    \end{pmatrix},
\end{multline}
which precisely yields the transformation $U D^{\rm QE}(S) U^\dagger = D^{\rm optimal}(S) \equiv D^{\rm ideal}(S)$, as illustrated in the bottom row of Fig.~\ref{fig:spinful}.

The model resulting from the considerations above read as
\begin{multline}
    H_{\rm sfg} = 
    \begin{pmatrix}
        c_0  & 0 & -c_2 k_- & 0
        \\
        0 & c_0 & 0 & -c_2 k_+
        \\
        -c_2 k_+ & 0 & c_1 & 0
        \\
        0 & -c_2 k_- & 0 & c_1
    \end{pmatrix}
    \\
    +
    \begin{pmatrix}
        c_4 k^2 & 0 & -c_5 k_+^2 & 0
        \\
        0 & c_4 k^2 & 0 & -c_5 k_-^2
        \\
        -c_5 k_-^2 & 0 & c_6 k^2 & 0
        \\
        0 & -c_5 k_+^2 & 0 & c_6 k^2
    \end{pmatrix},
    \label{eq:HgrapheneSOC}
\end{multline}
where $k^2 = k_x^2 + k_y^2$, $k_\pm = k_x \pm i k_y$, and we omit $k_z$-dependent for 2D materials. Notice that if we do not consider the composed magnetic anti-unitary symmetry $\mathcal{TI}$, the $c_2$ and $c_5$ terms above split into real and imaginary parts. Particularly for $c_2$, the real part refers to matrix elements of $\bm{p}$, while the imaginary part would carry contributions from $\bm{p}_{\rm soc}$. Nevertheless, considering $\mathcal{TI}$, these coefficients are expected to be real and the $\bm{p}_{\rm soc}$ contributions to the imaginary part vanish by symmetry.

The numerical values found for the parameters of $H_{\rm sfg}$ in Eq.~\eqref{eq:HgrapheneSOC} are shown in Table \ref{tab:graphene-soc}. The Fermi velocity matches the one from spinless graphene above, and we find that the intrinsic spin-orbit coupling is $\lambda_I = c_1-c_0 \approx 1$~$\mu$eV, which is much smaller than its established value of $\lambda_I \approx 24$~$\mu$eV obtained via all-electron full-potential DFT implementations \citep{graphene_soc, Avsar2020Colloquium}. This discrepancy is due to limitations of the pseudo-potentials used here with QE \cite{repository-ONCV}, which do not include d orbitals. Nevertheless, this example serves to show that, whenever two irreps are nearly degenerate, the DFT wavefunctions might always be mixed into reducible representations and the symmetry optimization procedure implemented here efficiently rotates the DFT basis back into ideal form that yields block-diagonal reducible representations.

\begin{table}[H]
    \renewcommand*{\arraystretch}{1.5}
    \caption{\label{tab:graphene-soc}Spinful graphene parameters for the Hamiltonian of Eq.~\eqref{eq:HgrapheneSOC}.}
    \begin{ruledtabular}
    \begin{tabular}{c|c|c}
    Coefficient & Values in a.u. & Values in (eV, nm)\\
    \hline
    $c_{0}$ & $-1.39\times 10^{-5}$ & $-0.000189$~eV\\
    $c_{1}$ & $-1.40\times 10^{-5}$ & $-0.000190$~eV\\
    $c_{2}$ & $0.72$ & $0.518$ ${\rm eV\,nm}$\\
    $c_{4}$ & $0.049$ & $0.0018$ ${\rm eV\,nm}^{2}$\\
    $c_{5}$ & $-0.82$ & -$0.031$ ${\rm eV\,nm}^{2}$\\
    $c_{6}$ & $0.049$ & $0.0018$ ${\rm eV\,nm}^{2}$\\
    \end{tabular}
    \end{ruledtabular}
\end{table}

\section{Examples}
\label{sec:Examples}

In this section, we briefly show the results for a series of selected
materials without presenting a step-by-step tutorial as above. More
details for each case below can be seen in the code repository. Here we consider
examples of zincblende crystals (GaAs, HgTe, CdTe), wurtzite crystals
(GaN, GaP, InP), rock-salt crystals (SnTe, PbSe), 
a transition metal dichalcogenide monolayer (${\rm MoS}_{2}$),
3D and 2D topological insulators (${\rm Bi}_{2}{\rm Se}_{3}$, ${\rm GaBiCl}_{2}$).
Additional examples can be found in the code repository.
In all cases, the resulting models agree well with the DFT bands near
the $\bm{k}\cdot\bm{p}$ expansion point and low energies, as expected.
The DFT parameters used in the simulations are presented in Appendix \ref{sec:DFT}.

\subsection{Zincblende crystals}

We consider well-known zincblende crystals: GaAs, CdTe and HgTe. These
crystals are characterized by lattices that transform as the space
group ${\rm F}\bar{4}3{\rm m}$, but their low energy bandstructure
concentrates near the $\Gamma$ point, which can be described by the
point group $T_{d}$ after factorizing the invariant subgroup of Bloch
translations. The basis functions and effective Kane model for these
materials are well described in the literature \cite{Winkler2003,Dresselhaus2007,Voon2009}.
Here, let us simply summarize this characterization to establish a
notation. 

In all cases considered in this section, the first conduction band
and the top valence bands transform either as $S$ or $P=(X,Y,Z)$
orbitals, and in terms of the crystallographic coordinates we define
$x\parallel[100]$, $y\parallel[010]$, and $z\parallel[001]$. In
the single group $T_{d}$, neglecting spin, the S-like orbitals transform
accordingly to the trivial $A_{1}$ irrep of $T_{d}$, while the P-like
orbitals transform as the $T_{2}$ irrep. Including spin, the double
group representation for the S-like orbitals become $A_{1}\otimes D_{1/2}=\bar{\Gamma}_{6}$,
where $D_{1/2}$ is the spinor representation, and it yields the spin
1/2 basis functions $\ket{S\uparrow}$ and $\ket{S\downarrow}$. For
the P-like bands one gets $T_{2}\otimes D_{1/2}=\bar{\Gamma}_{8}\oplus\bar{\Gamma}_{7}$,
where $\bar{\Gamma}_{8}$ represents the basis functions of total
angular momentum 3/2, and $\bar{\Gamma}_{7}$ has total angular momentum
1/2. These basis functions are listed in Table \ref{tab:BasisZB}.
For GaAs and CdTe the conduction band is represented by $\bar{\Gamma}_{6}$
(S-type, and spin 1/2), the first valence band is composed of P-type
orbitals with total angular momentum 3/2, which are described by the
$\bar{\Gamma}_{8}$ irrep, and the split-off band contains P-type
orbitals with total angular momentum 1/2, which defines the irrep
$\bar{\Gamma}_{7}$. In contrast, for HgTe the $\bar{\Gamma}_{6}$
and $\bar{\Gamma}_{8}$ are inverted due to fine structure corrections.

The basis from Table \ref{tab:BasisZB} diagonalizes the spinful effective
Hamiltonian at $\bm{k}=0$, and leads to the well known extended
Kane Hamiltonian \cite{Winkler2003}. The expression for the $8\times8$
Hamiltonian $H_{{\rm ZB}}$ is shown in Appendix \ref{app:HandCs}
in terms of the coefficients $c_{j}$ following the output
of the \texttt{qsymm} code, so that it matches \texttt{Examples} in
our repository. There, the notation for the powers of $\bm{k}$ follows
from Ref. \cite{Winkler2003}, such that it can be directly compared
to the extended Kane model shown in their Appendix C. The values for
the coefficients $c_{j}$ are also shown in Appendix \ref{app:HandCs}.

\begin{figure*}[th]
    \includegraphics[width=2\columnwidth]{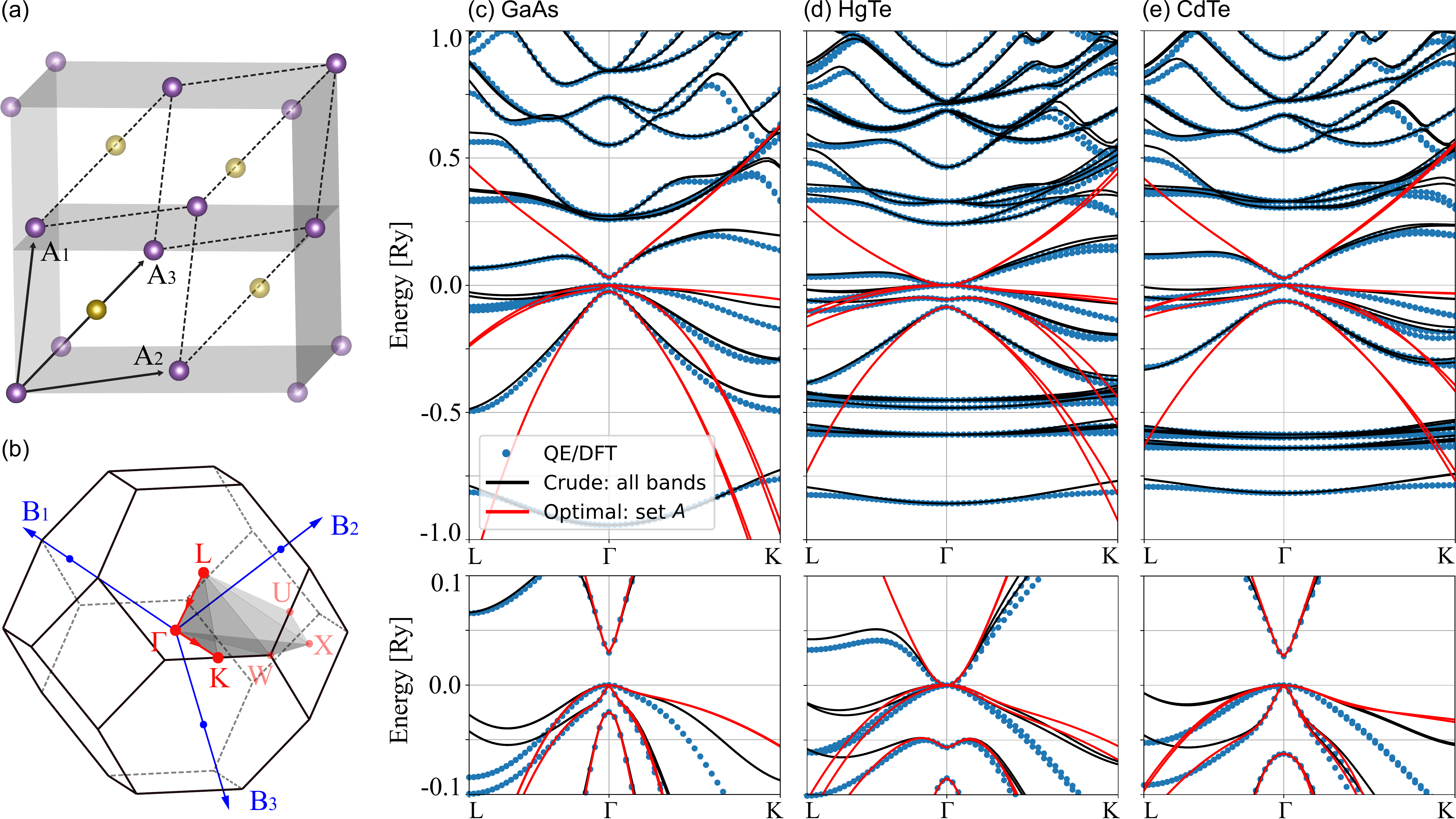}
    \caption{\label{fig:zincblende}(a) Zincblende lattice, and (b) its first Brillouin
    zone (FCC). The band structure for (c) GaAs, (d) HgTe, and (e) CdTe
    are shown over a large energy scale on the main panels, while at the
    bottom of each panel, we show a zoom over the relevant low energy range.
    In all cases, the DFT data consider 1000 bands.}
\end{figure*}

\newpage
The band structures calculated from $H_{{\rm ZB}}$ are shown in Fig.
\ref{fig:zincblende}, which also shows the crystal lattice and the
first Brillouin zone in Figs. \ref{fig:zincblende}(a-b).
In all cases, Figs. \ref{fig:zincblende}(c--e), the blue dots represent
the DFT results. The black lines are the crude model from Eq.
\ref{eq:crude_all_bands}, which includes all DFT bands and approaches
a full zone description, but with a cost of a large $N\times N$ model
with typical $N\gg100$. More importantly, the red lines represent
effective $8\times8$ Kane model from $H_{{\rm ZB}}$, which matches
well the DFT data at low energies and near $\Gamma$, as shown in
the zoomed insets below each panel for GaAs [Fig.~\ref{fig:zincblende}(c)],
HgTe [Fig.~\ref{fig:zincblende}(c)], and CdTe [Fig.~\ref{fig:zincblende}(c)].
Particularly, for HgTe it is clear the band inversion between the
$\bar{\Gamma}_{6}$ and $\bar{\Gamma}_{8}$ irreps.

\begin{table}[H]
    \renewcommand*{\arraystretch}{2.0}
    \caption{\label{tab:BasisZB}Basis functions for zincblende crystals. The first
    column indicates the double group irreps for the $T_{d}$ point group
    at $\Gamma$, which are induced from the single group irreps in parenthesis.
    The second column lists the basis functions on the basis of total
    angular momentum, and the third column shows their expressions in terms
    of the symmetry orbitals (S, X, Y, Z) and spin ($\uparrow$, $\downarrow$),
    which follows the definitions from Ref. \cite{Winkler2003}.}
    \begin{ruledtabular}
    \begin{tabular}{c|c|c}
    IRREP $T_{d}$ & $\ket{J,m_{j}}$ & $\ket{{\rm orb},{\rm spin}}$\\
    \hline 
    \multirow{2}{*}{$\bar{\Gamma}_{6}(A_{1})$} & $\ket{\frac{1}{2},+\frac{1}{2}}$ & $\ket{S,\uparrow}$\\
    \cline{2-3} \cline{3-3} 
     & $\ket{\frac{1}{2},-\frac{1}{2}}$ & $\ket{S,\downarrow}$\\
    \hline 
    \multirow{4}{*}{$\bar{\Gamma}_{8}(T_{2})$} & $\ket{\frac{3}{2},+\frac{3}{2}}$ & $-\frac{1}{\sqrt{2}}\ket{X+iY,\uparrow}$\\
    \cline{2-3} \cline{3-3} 
     & $\ket{\frac{3}{2},-\frac{3}{2}}$ & $+\frac{1}{\sqrt{2}}\ket{X-iY,\downarrow}$\\
    \cline{2-3} \cline{3-3} 
     & $\ket{\frac{3}{2},-\frac{1}{2}}$ & $+\frac{1}{\sqrt{6}}\Big[2\ket{Z,\downarrow}+\ket{X-iY,\uparrow}\Big]$\\
    \cline{2-3} \cline{3-3} 
     & $\ket{\frac{3}{2},+\frac{1}{2}}$ & $+\frac{1}{\sqrt{6}}\Big[2\ket{Z,\uparrow}-\ket{X+iY,\downarrow}\Big]$\\
    \hline 
    \multirow{2}{*}{$\bar{\Gamma}_{7}(T_{2})$} & $\ket{\frac{1}{2},-\frac{1}{2}}$ & $+\frac{1}{\sqrt{3}}\Big[\ket{Z,\downarrow}-\ket{X-iY,\uparrow}\Big]$\\
    \cline{2-3} \cline{3-3} 
     & $\ket{\frac{1}{2},+\frac{1}{2}}$ & $-\frac{1}{\sqrt{3}}\Big[\ket{Z,\uparrow}+\ket{X+iY,\downarrow}\Big]$\\
    \end{tabular}
    \end{ruledtabular}
\end{table}

\subsection{Wurtzite crystals}

\begin{figure*}[th]
    \includegraphics[width=2\columnwidth]{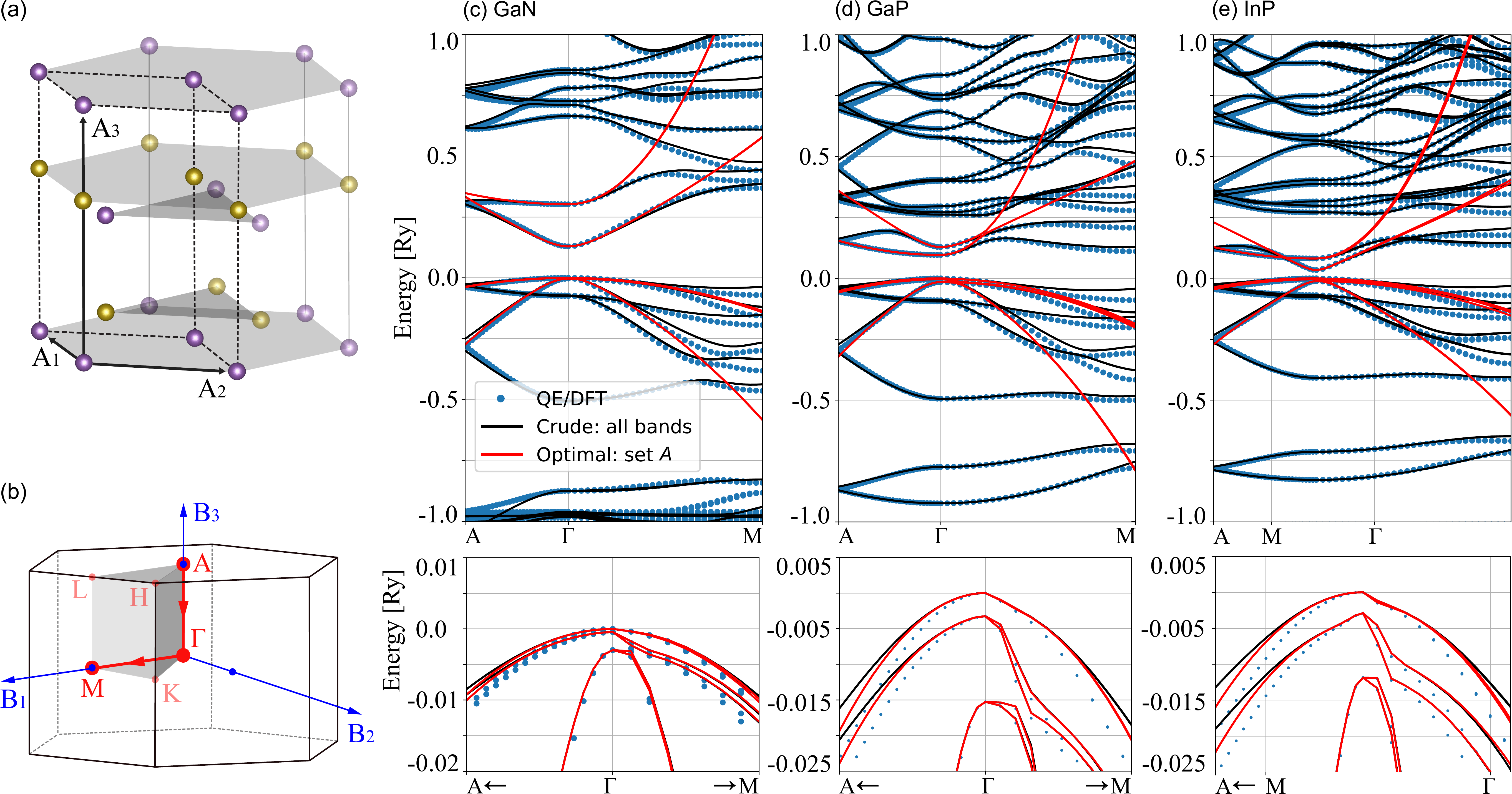}
    \caption{\label{fig:wurtzite}(a) Lattice and (b) Brillouin zone for wurtzite
    crystals. Band structures for (c) GaN, (d) GaP, and (e) InP show the large
    energy range on top, and a zoom shows the top of the valence bands
    at the bottom of each panel. In all cases, the DFT calculation considers
    1000 bands.}
\end{figure*}

The wurtzite crystals form a lattice that is characterized by the
space group ${\rm P}6_{3}{\rm mc}$, and the low energy band structure
appears near the $\Gamma$ point only. Near $\Gamma$, one can factorize
the translations and the resulting factor group is the $C_{6V}$ point
group, which is generated by the $C_{6}$ rotation around the z-axis,
and the mirror $M_{x}$. Here, in terms of the crystallographic coordinates,
$x\parallel[100]$, $y\parallel[010]$, and $z\parallel[001]$. The
unit cell and first Brillouin zone for these materials are shown in
Figs. \ref{fig:wurtzite}(a) and \ref{fig:wurtzite}(b).

To illustrate the results for wurtzite materials, we consider the
cases of GaN, GaP, and InP. Their band structures are shown in Figs.
\ref{fig:wurtzite}(c--e). In all cases, the top valence bands are
characterized by the irreps $(A_{1}+E_{1})\otimes D_{1/2}=\bar{\Gamma}_{7}\oplus2\bar{\Gamma}_{9}$.
Here, $A_{1}$ is the trivial irrep of $C_{6V}$ (single group), which
represents S-like and Z-like orbitals, and $E_{1}$ is the vector
representation of $C_{6V}$ that contains (X, Y)-like orbitals. These
are composed with the pure spinor representation $D_{1/2}$ to define
the $C_{6V}$ double group irreps $\bar{\Gamma}_{7}$ and $\bar{\Gamma}_{9}$.
Additionally, we consider two conduction bands, which are characterized
by the irreps $(A_{1}+B_{1})\otimes D_{1/2}=\bar{\Gamma}_{8}\oplus\bar{\Gamma}_{9}$.
The orbital basis function for the $B_{1}$ irrep is odd under both
$C_{6}$ and $M_{x}$, its representation on group character tables
is cumbersome, so one defines it as $\ket{X(X^{2}-3Y^{2})}\equiv\ket{V}$
\cite{Voon2009}. Ultimately, we consider the double group representations
ordered as shown in Table \ref{tab:BasisWZ}.

\begin{table}[H]
    \renewcommand*{\arraystretch}{2.0}
    \caption{\label{tab:BasisWZ}Basis functions for wurtzite crystals. The first column shows the double group irreps of $C_{6V}$, which are induced from the single group irrep between parenthesis. The second column shows the basis representation in terms of the spherical harmonics $Y_{l}^{m}$ and spin ($\uparrow$, $\downarrow$), while the third column shows the representation in terms of the orbitals (S, X, Y, Z, V), where $V=X(X^{2}-3Y^{2})$ \cite{Voon2009}.}
    \begin{ruledtabular}
    \begin{tabular}{c|c|c}
    IRREP $C_{6V}$ & $\ket{Y_{l}^{m},{\rm spin}}$ & $\ket{{\rm orb},{\rm spin}}$\\
    \hline 
    \multirow{2}{*}{$\bar{\Gamma}_{9}^{c}(A_{1})$} & $\ket{Y_{0}^{0},\uparrow}$ & $\ket{S',\uparrow}$\\
    \cline{2-3} \cline{3-3} 
     & $\ket{Y_{0}^{0},\downarrow}$ & $\ket{S',\downarrow}$\\
    \hline 
    \multirow{2}{*}{$\bar{\Gamma}_{8}^{c}(B_{1})$} & $\ket{Y_{3}^{3}-Y_{3}^{-3},\uparrow}$ & $\ket{V,\uparrow}$\\
    \cline{2-3} \cline{3-3} 
     & $\ket{Y_{3}^{3}-Y_{3}^{-3},\downarrow}$ & $\ket{V,\downarrow}$\\
    \hline 
    \multirow{2}{*}{$\bar{\Gamma}_{9}^{v}(A_{1})$} & $\ket{Y_{1}^{0},\uparrow}$ & $\ket{Z',\uparrow}$\\
    \cline{2-3} \cline{3-3} 
     & $\ket{Y_{1}^{0},\downarrow}$ & $\ket{Z',\downarrow}$\\
    \hline 
    \multirow{2}{*}{$\bar{\Gamma}_{9}^{v}(E_{1})$} & $\ket{Y_{1}^{1},\uparrow}$ & $\ket{X'+iY',\uparrow}$\\
    \cline{2-3} \cline{3-3} 
     & $\ket{Y_{1}^{-1},\downarrow}$ & $\ket{X'-iY',\downarrow}$\\
    \hline 
    \multirow{2}{*}{$\bar{\Gamma}_{7}^{v}(E_{1})$} & $\ket{Y_{1}^{-1},\uparrow}$ & $\ket{X'-iY',\uparrow}$\\
    \cline{2-3} \cline{3-3} 
     & $\ket{Y_{1}^{1},\downarrow}$ & $\ket{X'+iY',\downarrow}$\\
    \end{tabular}
    \end{ruledtabular}
\end{table}

\newpage
There the top indexes
$\{c,v\}$ refer to conduction and valence bands. Notice that the
$\Gamma_{9}$ irrep appears in three pairs of basis functions, which
allows for the $s$--$p_{z}$ mixing 
\cite{rashba1959symmetry, LewYanVoon1996, litvinov2016wide}
Here, however, we always work on the diagonal basis ($H_{{\rm WZ}}$ is
diagonal at $\bm{k}=0$), which is indicated by the primes in the
orbitals above. For a recent and detailed discussion on this choice
of representation and the $s$--$p_z$ mixing, please refer to Ref. \cite{Fu2020}.

Using the basis functions from Table \ref{tab:BasisWZ} to calculate
the effective $10\times10$ model using \texttt{qsymm}, we obtain
the Hamiltonian $H_{{\rm WZ}}$ shown in Appendix \ref{app:HandCs}.
Here we always consider two conduction bands, which leads to this
$10\times10$ generic model $H_{{\rm WZ}}$. However, one can also
opt to work with traditional $8\times8$ models with a single conduction
band. Notice, however, that for GaP the first conduction band transforms
as $\bar{\Gamma}_{8}$, while for GaN and InP the first conduction
band is $\bar{\Gamma}_{9}$. Therefore, one must be careful when selecting
the appropriate $8\times8$ model for wurtzite materials. For the
valence bands, one always gets $\bar{\Gamma}_{7}\oplus2\bar{\Gamma}_{9}$,
however, the internal ordering of these valence bands may change between
materials and it can be highly sensible to the choice of density functional
\cite{Paulo2016, Campos2018, daSilva2020, Bonani2021}. 
The numerical coefficients $c_{j}$ found for GaN, GaP,
InP are shown in Appendix \ref{app:HandCs}, and the resulting
band structures are shown in Figs. \ref{fig:wurtzite}(c--e). 
In all cases, we see that the crude model with 1000 bands (black
lines) approaches a full zone description, but here we are more interested
in the reduced $10\times10$ models (red lines), which present satisfactory
agreement with the DFT data at low energies.

\subsubsection{Effects of the SOC corrections on \texorpdfstring{$\bm{P}_{m,n}$}{Pmn}}
\label{sec:PSOC}

As introduced in Sec. \ref{sec:Pmn}, the matrix elements $\bm{P}_{m,n}$ can be calculated with or without the PAW corrections, $\bm{p}_{\rm SOC}$, that carry the SOC contributions. For most of the materials we have studied here, these corrections are marginal and the results from both cases are nearly identical. Nevertheless, we emphasize that using our patched \texttt{bands.x} within QE is faster than using the Python code to calculate $\bm{P}_{m,n}$ via Eq.~\eqref{eq:PmnNoPAW}.

\begin{figure}[th]
    \includegraphics[width=\columnwidth]{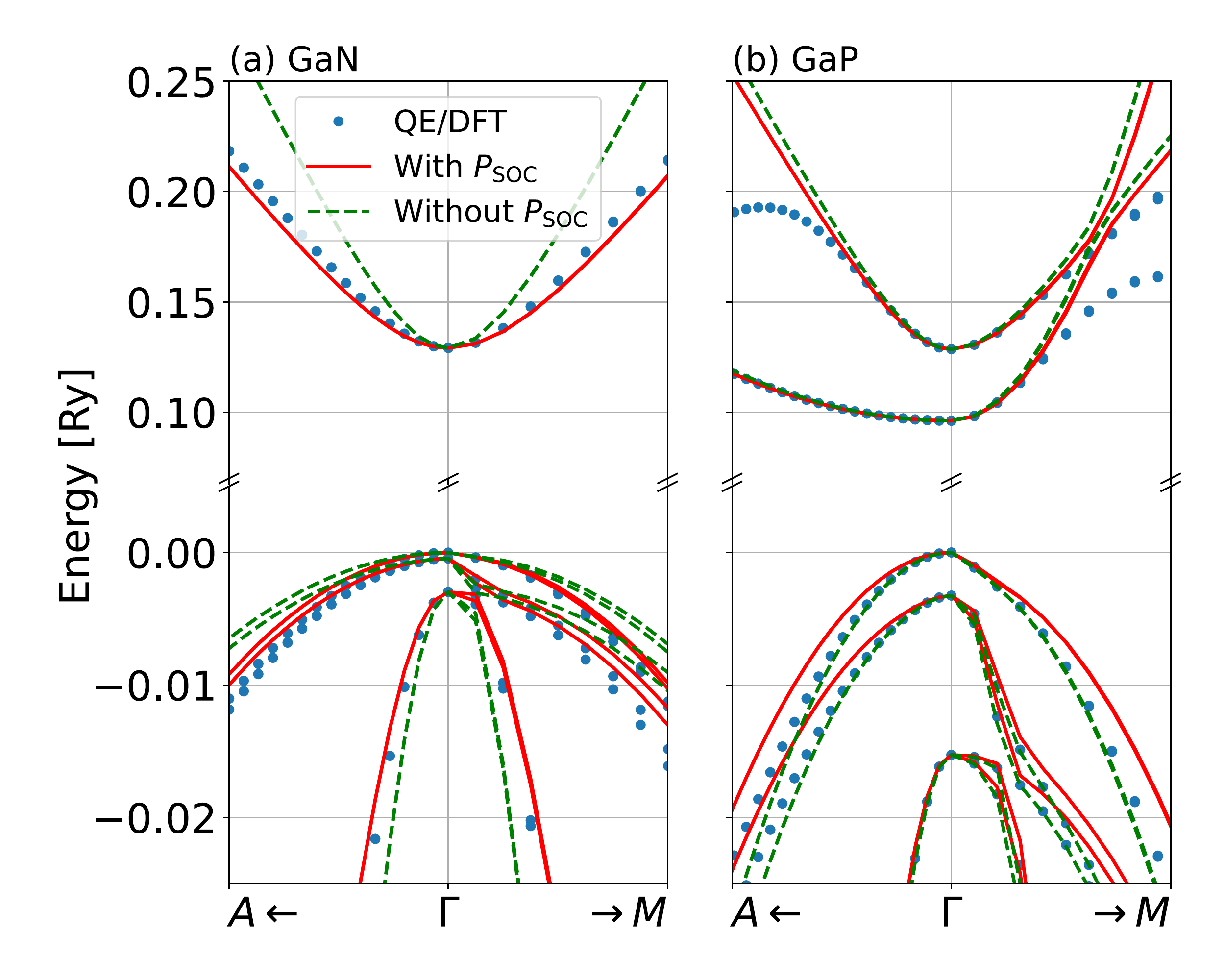}
    \caption{\label{fig:GaNGaPSOC} 
    Comparison between the DFT data and the effective models calculated with the full matrix element $\bm{P}_{m,n}$ including PAW/SOC corrections (red lines) and the simplified $\bm{P}_{m,n}$ without PAW/SOC corrections (green lines) for (a) GaN and (b) GaP.}
\end{figure}

To illustrate the effects of the PAW/SOC corrections on the matrix elements $\bm{P}_{m,n}$, Fig.~\ref{fig:GaNGaPSOC} compares the models for GaN and GaP with and without these corrections. For the conduction bands, we notice that the $\bm{p}_{\rm SOC}$ corrections significantly improve the GaN effective mass, but barely affect GaP. For the valence bands, both GaN and GaP show moderate effects of $\bm{p}_{\rm SOC}$. Indeed, this shows that a precise calculation of $\bm{P}_{m,n}$ is critical to improve the precision of the models \cite{Psoc}.

\newpage
\subsection{Rock-salt crystals}

\begin{figure}[th]
    \includegraphics[width=1\columnwidth]{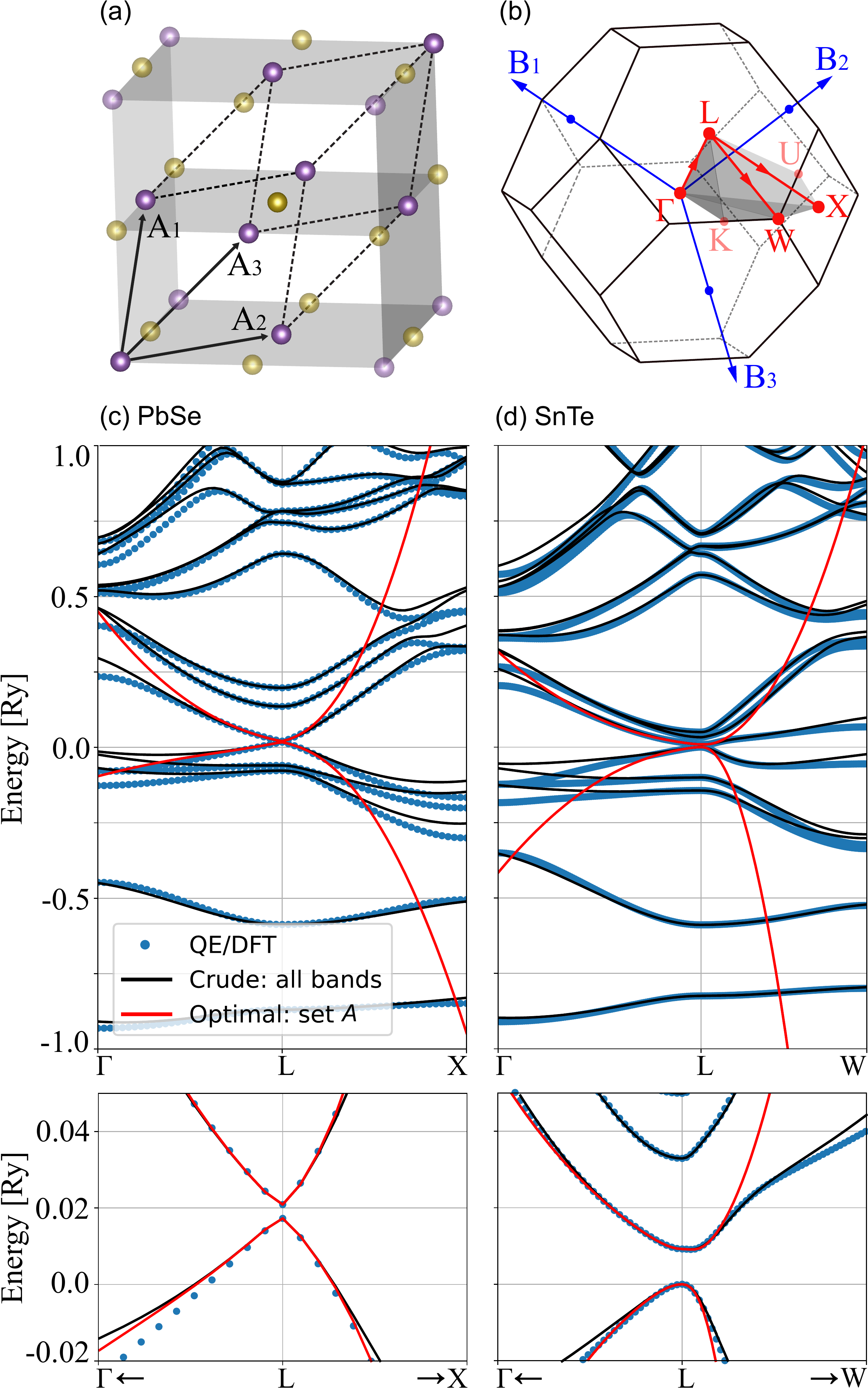}
    \caption{\label{fig:rocksalt}(a) The rock salt lattice and (b) its Brillouin
    zone (FCC). Band structures for (c) PbSe and (d) SnTe. The bottom
    of each panel zooms into the low energy range near the Fermi level.
    Both DFT calculations were performed considering 500 bands.}
\end{figure}

The crystal lattice for rock-salt crystals is shown in Fig.~\ref{fig:rocksalt}(a),
which is an FCC lattice with two atoms in the base, and it is described
by the space group ${\rm Fm}\bar{3}{\rm m}$. The low energy band
structure concentrates at the L point of the Brillouin zone shown
in Fig.~\ref{fig:rocksalt}(b), which transforms as the $D_{3D}$
point group after factorizing the Bloch translations. The basis functions
for the first valence and conduction bands transform as $A_{1g}\otimes D_{1/2}=\bar{L}_{6}^{+}$
and $A_{2u}\otimes D_{1/2}=\bar{L}_{6}^{-}$, where $A_{1g}$ is the
trivial irrep for S-like orbitals, and $A_{2u}$ represent Z-like
orbitals \cite{Mitchell1966}. Therefore, the basis functions for
the $\bar{L}_{6}^{+}$ bands are $\{\ket{S,\uparrow},\ket{S,\downarrow}\}$,
and for $\bar{L}_{6}^{-}$ one gets $\{\ket{Z,\uparrow},\ket{Z,\downarrow}\}$.
Here, the $x$, $y$, and $z$ coordinates are taken along the $[\bar{1}\bar{1}2]$,
$[1\bar{1}0]$, and $[111]$ crystallographic directions.

Here we consider two examples of rock-salt crystals: PbSe and SnTe.
Their effective $4\times4$ Hamiltonian $H_{{\rm RS}}$ under the
$\bar{L}_{6}^{\pm}$ basis, and its numerical parameters are shown
in Appendix \ref{app:HandCs}, and the comparison between
DFT and model band structures are shown in Figs. \ref{fig:rocksalt}(c)--(d).
PbSe is a narrow gap semiconductor, where the conduction band transforms as 
the $\bar{L}_6^+$ irrep, and the valence band as $\bar{L}_6^-$. 
In contrast, SnTe shows inverted bands, with $\bar{L}_6^+$ below $\bar{L}_6^-$,
yielding a topological insulator phase \cite{SnTe_TCI_1,SnTe_TCI_2}.
In both cases, the low-energy model captures the main features of the
bands, including the anisotropy.

\newpage
\subsection{Other examples}

To finish the set of illustrative examples, we show here the case for:
(i) the monolayer ${\rm MoS}_{2}$, which is one of the most studied 
transition metal dichalcogenides (TMDC) \cite{MoS2_1,MoSe_2,MoSe_3};
(ii) the bulk bismuth selenide (${\rm Bi}_{2}{\rm Se}_{3}$),
which is one of the first discovered 3D topological insulators \cite{Bi2Se3_TI_1,Bi2Se3_TI_2};
and (iii) a monolayer of ${\rm GaBiCl}_{2}$, which is
a large gap 2D topological insulator \cite{Li2015}. The symmetry characteristics and basis functions for the low-energy
bands of these materials mentioned above are summarized in Table
\ref{tab:2DSummary}.

\begin{table}[H]
    \renewcommand*{\arraystretch}{2.0}
    \caption{\label{tab:2DSummary}Summary of space group, irreps and basis functions
    for the low energy bands of ${\rm MoS}_{2}$, ${\rm GaBiCl}_{2}$, and ${\rm Bi}_{2}{\rm Se}_{3}$.
    The first column lists
    the materials, the second indicates the lattice space group, and the
    little group at the relevant k point. The third and fourth columns
    lists the irreps and basis functions for the low energy bands in each
    case. The table shows the double group irreps and the corresponding
    single group irreps between parenthesis.}
    \begin{ruledtabular}
    \begin{tabular}{c|c|c|c}
    Material & Group info & IRREP & Basis\\
    \hline 
    \multirow{4}{*}{${\rm MoS}_{2}$} & Space group & $\bar{K}_{11}(E_{1}')$ & $\ket{X+iY,\uparrow}$\\
    \cline{2-4} \cline{3-4} \cline{4-4} 
     & ${\rm P}\bar{6}m2$ & $\bar{K}_{10}(E_{1}')$ & $\ket{X+iY,\downarrow}$\\
    \cline{2-4} \cline{3-4} \cline{4-4} 
     & Little group & $\bar{K}_{8}(A')$ & $\ket{S,\uparrow}$\\
    \cline{2-4} \cline{3-4} \cline{4-4} 
     & K: $C_{3h}$ & $\bar{K}_{9}(A')$ & $\ket{S,\downarrow}$\\
    \hline 
    \hline 
    \multirow{6}{*}{${\rm GaBiCl}_{2}$} & \multirow{2}{*}{Space group} & $\bar{\Gamma}_{4}(E)$ & $\ket{X+iY,\uparrow}$\\
    \cline{3-4} \cline{4-4} 
     &  & $\bar{\Gamma}_{5}(E)$ & $\ket{X-iY,\downarrow}$\\
    \cline{2-4} \cline{3-4} \cline{4-4} 
     & ${\rm P3m1}$ & \multirow{2}{*}{$\bar{\Gamma}_{6}(E)$} & $\ket{X-iY,\uparrow}$\\
    \cline{2-2} \cline{4-4} 
     & \multirow{2}{*}{Little group} &  & $\ket{X+iY,\downarrow}$\\
    \cline{3-4} \cline{4-4} 
     &  & \multirow{2}{*}{$\bar{\Gamma}_{6}(A_{1})$} & $\ket{Z\uparrow}$\\
    \cline{2-2} \cline{4-4} 
     & $\Gamma$: $C_{3V}$ &  & $\ket{Z\downarrow}$\\
    \hline 
    \hline 
    \multirow{4}{*}{${\rm Bi}_{2}{\rm Se}_{3}$} & Space group & \multirow{2}{*}{$\bar{\Gamma}_{6}^{+}(A_{1g})$} & $\ket{S,\uparrow}$\\
    \cline{2-2} \cline{4-4} 
     & ${\rm R}\bar{3}{\rm m}$ &  & $\ket{S,\downarrow}$\\
    \cline{2-4} \cline{3-4} \cline{4-4} 
     & Little group & \multirow{2}{*}{$\bar{\Gamma}_{6}^{-}(A_{2u})$} & $\ket{Z,\uparrow}$\\
    \cline{2-2} \cline{4-4} 
     & $\Gamma$: $D_{3d}$ &  & $\ket{Z,\downarrow}$\\
    \end{tabular}
    \end{ruledtabular}
\end{table}

For ${\rm MoS}_{2}$, the first valence and conduction bands are given
by the single group irreps $A'$ and $E'_{1}$ of the $C_{3h}$ group
\cite{Kormanyos2015,Rybkovskiy2017}, which can be represented as
S-like and $(X+iY)$-like orbitals.
For ${\rm GaBiCl}_{2}$, the valence bands are characterized by single
group $E$ irrep, and it splits into $E\otimes D_{1/2}=\bar{\Gamma}_{4}\oplus\bar{\Gamma}_{5}\oplus\bar{\Gamma}_{6}$
in the spinful case, while the conduction band is given by the irrep
$A_{1}\otimes D_{1/2}=\bar{\Gamma}_{6}$. 
For ${\rm Bi}_{2}{\rm Se}_{3}$,
a detailed derivation of the effective model can be seen in Ref. \cite{Liu2010},
which shows that the first valence and conduction bands are given
by $A_{1g}\otimes D_{1/2}=\Gamma_{6}^{+}$, and $A_{2u}\otimes D_{1/2}=\Gamma_{6}^{-}$.

\begin{figure*}[th]
    \includegraphics[width=2\columnwidth]{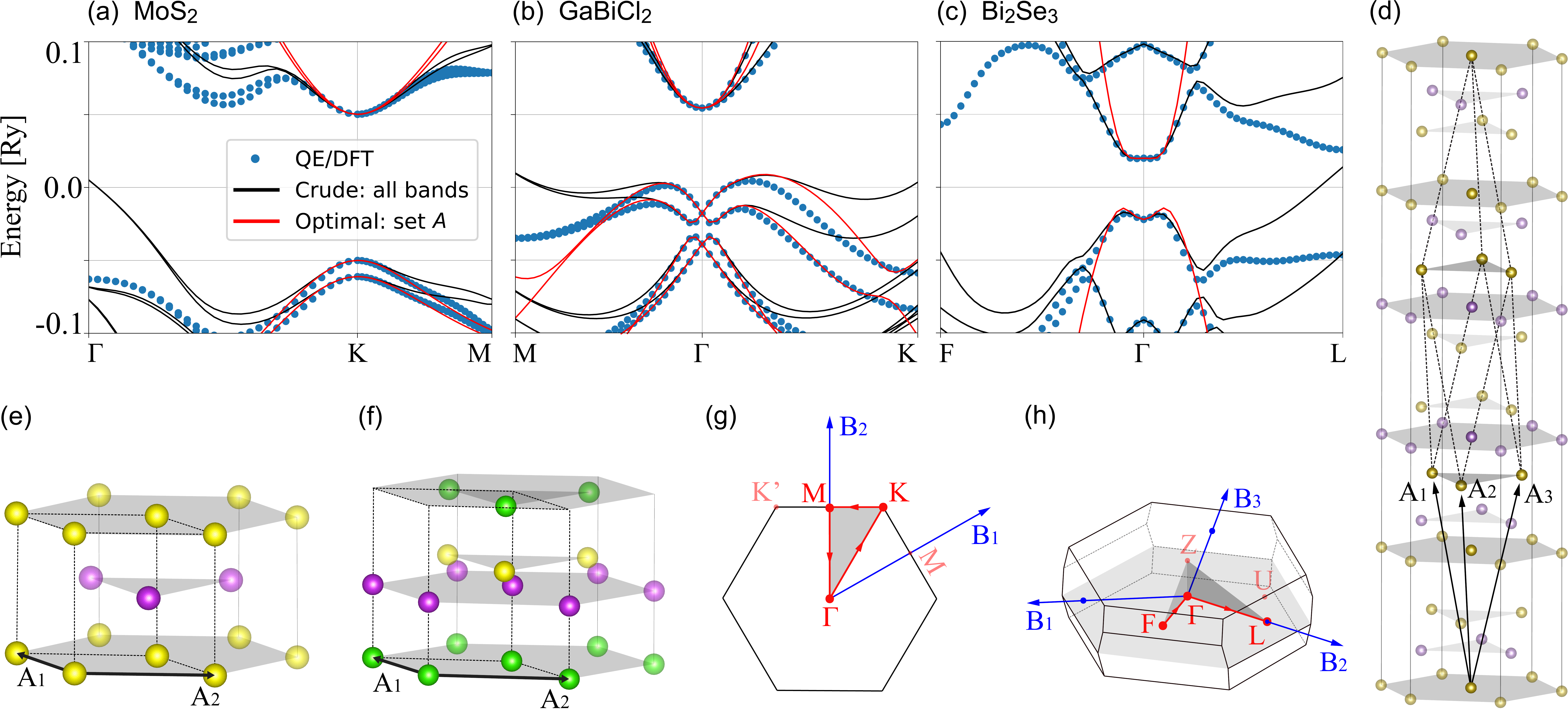}
    \caption{\label{fig:OtherMaterials}Band structures for: 
    (a) ${\rm MoS}_{2}$, (b) ${\rm GaBiCl}_{2}$, and (c) ${\rm Bi}_{2}{\rm Se}_{3}$ showing only the relevant low energy range. The DFT calculations were performed for 1000, 500, and 500 bands, respectively. (d) Rhombohedral lattice of ${\rm Bi}_{2}{\rm Se}_{3}$ and 2D hexagonal lattice of (e) ${\rm MoS}_{2}$ and (f) ${\rm GaBiCl}_{2}$, where we have omitted the vacuum region (15 Å) perpendicular to the plane formed by vectors A$_1$ and A$_2$. (g) 2D Brillouin zone common to ${\rm MoS}_{2}$ and ${\rm GaBiCl}_{2}$, and (h) 3D BZ of ${\rm Bi}_{2}{\rm Se}_{3}$.}
\end{figure*}

The effective Hamiltonians and their numerical coefficients for these
materials can be found in the \texttt{Examples} folder of the code repository.
Here we show only the comparison between the DFT and model band structures 
in Fig.~\ref{fig:OtherMaterials}. 
The ${\rm MoS}_{2}$ case, as shown in Fig.~\ref{fig:OtherMaterials}(a),
is challenging for a $\bm{k}\cdot\bm{p}$ method, since its band structure presents valleys
in between high symmetry points. Consequently, the 4 bands
model (red lines) captures only the nearly parabolic dispersion at
the K point.

However, the crude all-bands model (black lines,
see Eq.~\eqref{eq:crude_all_bands}) approaches a full zone description
and captures the valley along the $\Gamma$--K direction. 
For ${\rm GaBiCl}_{2}$, Fig.~\ref{fig:OtherMaterials}(c), the 6 bands model describes satisfactorily the low energy conduction and
valence bands.
For ${\rm Bi}_{2}{\rm Se}_{3}$ in Fig.~\ref{fig:OtherMaterials}(b) the
4 bands model captures well the low-energy band structure
near $\Gamma$, including the hybridization between the inverted bands.

\section{Discussions}
\label{sec:discussions}

Above, we have presented illustrative results of the capabilities
of our code to calculate the $\bm{k}\cdot\bm{p}$ Kane and Luttinger
parameters for a series of relevant materials. In all cases we see
a patent agreement between the DFT (QE) data and the low-energy
models near the relevant $\bm{k}_0$ point. However, it is important to notice
that here we use only PBE functionals \cite{GGA-PBE}, consequently it often
underestimates the gap (e.g. 0.5 eV instead of 1.5 eV for GaAs). Therefore,
our models are limited by the quality of the DFT bands and the resulting
numerical parameters might not match Kane and Luttinger's parameters
for well-known materials, for which these parameters are typically
chosen to match the experimental data, and not the DFT simulations.

For instance, let us consider the zincblende crystals' Kane parameter
$E_P = 2m_0P^2/\hbar^2$, band gap $E_g$ and effective mass for the conduction band $m^*$.
For GaAs, the experimental values are $E_P \sim 24$~eV, $P\sim 0.96$~eVnm,
$E_g \sim 1.5$~eV, and $m^* = 0.065 m_0$ \cite{Vurgaftman2001JAP}.
As mentioned above, the DFT results with PBE functionals underestimate the gap,
and we get $E_g \sim 0.5$~eV. Moreover, the Kane parameter can be written
as $P = -\sqrt{6} c_5/2$, where the coefficient $c_5 = -0.635$~eVnm is shown 
in Appendix \ref{app:HandCs}. This value yields $P \sim 0.7$~eVnm and $E_P \sim 16$~eV.
The effective mass for the conduction band can be estimated from its
spinless expression \cite{Kane1957}, $m_0/m^* = 1 + 2m_0P^2/E_g\hbar^2$,
which gives us $m^* = 0.031m_0$. While these numbers do not match well with the
experimental values, we notice that if we fix the GaAs gap (scissors-cut approximation),
but keep our value for $P$, we find $m^* = 0.058m_0$, which is already
much closer to the experimental value for the effective mass.

The number estimates shown above clearly indicate that the quality of our
models is limited to the DFT simulations only. Particularly, the gap issue can be 
fixed if one replaces the PBE functionals with hybrid functionals, GW calculations, 
or other methods that improve the material gap accuracy. 
These are beyond the scope of this paper, but it is a possible path for future
improvements of our code.

In all examples presented here, we always consider the crude
all bands model from Eq.~\eqref{eq:crude_all_bands}, and the optimal symmetry-adapted
(few bands) model from Eq.~\eqref{eq:lowdin}. This raises two interesting
questions: (i) how many bands are necessary for convergence? And (ii)
for a large number of bands, should we get a full zone description?
We discuss these questions below.

\newpage
\subsection{Convergence}

The convergence threshold (how many bands are necessary) strongly depends on the material. In some cases $\sim 300$ bands are sufficient, but in others, it often needs $\sim 1000$ bands. We do not have a general rule to establish which materials will show a slow or fast convergence. Nevertheless, we believe it is instructive to discuss the outcomes of our convergence analysis.

Notice that the L\"owdin partitioning from Eq.~\eqref{eq:lowdin} has two distinct contributions. The first two terms in Eq.~\eqref{eq:lowdin} are the zeroth and first-order perturbation terms. These terms do not change as we increase the number of DFT bands (provided that there are enough bands to converge the DFT calculation itself). The zeroth order term is essentially given by the DFT eigenstates, and the first order terms are given by the matrix elements $\bra{m}H'(\bm{k})\ket{n}=2\bm{k}\cdot\bm{P}_{m,n}$ between eigenstates of set $A$, which is the low energy sector of interest. In contrast, the third term defines the second-order corrections, which are quadratic in $\bm{k}$ (assuming a diagonal basis at $\bm{k}=0$). In this case, the second-order contributions depend explicitly on the sum over the remote set of bands $B$. These are the terms that strongly depend on the number of remote bands.

\begin{figure*}[th]
    \includegraphics[width=1\textwidth]{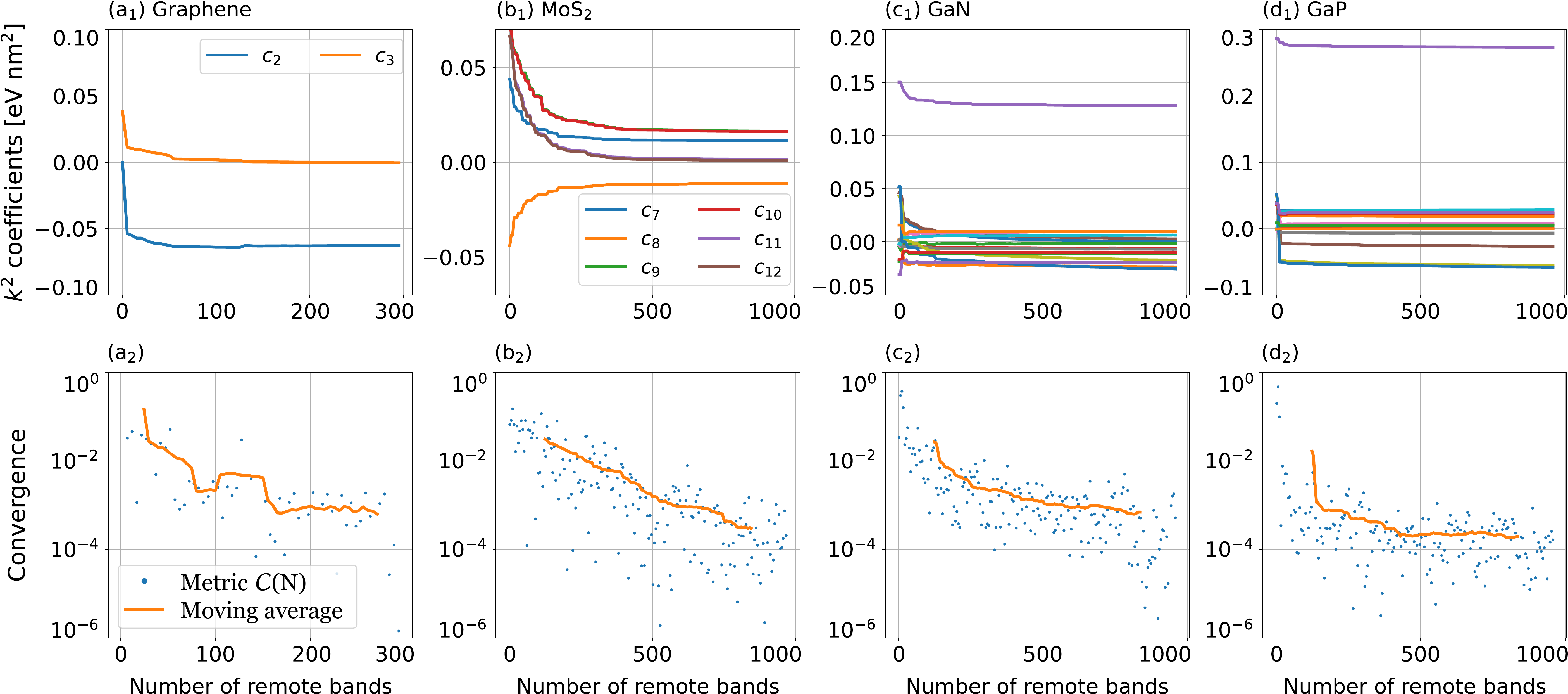}
    \caption{\label{fig:Convergence}
    Convergence of the second-order coefficients
    $c_{j}$ as a function of the number of remote bands for (a) spinless
    graphene, (b) ${\rm MoS}_{2}$, (c) GaN, and (d) GaP. On top (a$_1$--d$_1$), each panel shows the 
    coefficients $c_{j}$ for different material. On panels (c$_1$) and (d$_1$) we omit the 
    legends because there are 30 distinct coefficients, ranging from $c_{22}$ to $c_{51}$, 
    which makes their individual identification cumbersome, and it is sufficient to visualize 
    that all lines become nearly flat for a large number of remote bands.
    On the bottom (a$_2$--d$_2$), for each material, the evolution of the coefficients $c_j$
    are combined into convergence metric set by Eq.~\eqref{eq:metric} (blue dots).
    Due to the noise induced by the discrete derivative in this metric, 
    we plot the moving average of the data as a guide for the eyes.}
\end{figure*}

To check for convergence, we plot the values of the Hamiltonian coefficients $c_{j}$ associated with second-order corrections as a function of the number of remote bands. In the \texttt{Examples} folder in the code repository, one finds these plots for all cases presented in this paper. Here, in the top panels of Fig.~\ref{fig:Convergence}, we select a few illustrative cases. 
In the bottom panels of Fig.~\ref{fig:Convergence} we combine the discrete derivatives of $c_{j}$ into a single dimensionless metric for convergence $\mathcal{C}(N)$, which read as
\begin{align}
    \label{eq:metric}
    \mathcal{C}(N) &= \dfrac{\sum_j |c_j(N+1) - c_j(N)|}{\sum_j |c_j(N)|},
\end{align}
where $c_j(N)$ refers to the coefficient calculated using $N$ remote bands. With increasing $N$, the coefficients are expected to converge, consequently $\mathcal{C}(N) \rightarrow 0$. The data for $\mathcal{C}(N)$ is shown in blue dots on the bottom panels of Fig.~\ref{fig:Convergence}, which is significantly noisy due to the discrete jumps on the evolution of $c_j$ with increasing $N$. Therefore, we also plot a moving average $\mathcal{C}(N)$ (orange lines) to clearly show the convergence.
For spinless graphene in Fig.~\ref{fig:Convergence}(a), there are only two second order $c_{j}$ terms (neglecting terms with $k_z$, since it is a 2D material), and we see that it reaches convergence with less than 300 remote bands.

In contrast, for ${\rm MoS}_{2}$, the convergence requires at least $\sim500$ remote bands. Interestingly, it has been recently shown that TMDC materials indeed require a large number of bands to converge the orbital angular momenta \cite{Wozniak2020PRB, Deilmann2020PRL, Forste2020NatComm, Xuan2020PRR}. This fact may be associated with the large number of unoccupied bands with plane-wave character that appear due to the spatial extension of the vacuum region.
The GaN and GaP cases in Figs.~\ref{fig:Convergence}(c)--(d) are interesting cases, they belong to the same class of materials, but GaP reaches convergence with $\sim200$ remote bands, while GaN is not yet fully converged for $\sim1000$ remote bands. Unlike monolayer materials, the GaN compound is not described by any vacuum region, and therefore we speculate that such poor convergence may be related to details of the pseudopotential \cite{Kageshima1997} and the electronegativity of Nitrogen.

\subsection{Full zone kp}
\label{sec:fullzone}

In Section \ref{subsec:kp} we have presented the \kp method in its traditional form, which considers a perturbative expansion of the Bloch Hamiltonian at a reference momentum $\bm{k}_0$, and a small set of bands near the Fermi energy. Usually, one expects the resulting effective model to be valid only near $\bm{k}_0$ and only for a small energy range that encloses the bands of interest. In contrast, within the \textit{full zone} \kp approach \cite{Cardona1966, Radhia2002, Beresford2004, Saidi2008, Sadi2010, Gawarecki2022} one considers a large set of bands, such that the resulting low energy model agrees well with DFT or experimental bands over the full Brillouin zone, instead of only the vicinity of $\bm{k}_0$. However, to achieve this precision, one needs to apply fitting procedures to ensure that the bands match selected energy levels at various $\bm{k}$ points over the Brillouin zone. 

Here, in our code, we can easily select an arbitrary number of bands to build effective models. All examples presented above show sets of bands colored in red and black, such that the red ones consider models built from a small set of bands $A$ (from 4 to 10 bands), while the black ones consider the full set of bands from the DFT data (typically 500 or 1000 bands). This leads to an interesting question: should our \emph{all bands model} match the \emph{full zone} $\bm{k}\cdot\bm{p}$ models? 

To answer this question, let us focus first on the graphene results from Fig.~\ref{fig:graphene}. There, we have seen that the QE/DFT and the model agree remarkably well at low energies near the K point, as expected. Particularly, the red line for the optimal symmetry-adapted model describes precisely the low energy regime and Dirac cone and the trigonal warping from the quadratic terms in Eq.~\eqref{eq:Hgraphene}. In contrast, when we consider the all-bands model (black lines), we see that the model approaches a full zone agreement with 300 bands. What if we consider more bands? Our numerical tests have shown that increasing the number of bands does improve the overall description, approaching the full zone agreement. However, this is a very slow convergence and we never really reach a true full zone agreement. This characteristic is seen in all other examples shown here.

For GaAs, Gawarecki and collaborators \cite{Gawarecki2022} show an excellent full zone agreement between model and DFT bands considering 30 bands. In contrast, our results presented in Fig.~\ref{fig:zincblende}(a) for 8 (red) and 1000 (black) bands remain valid only in the vicinity of $\Gamma$. The key difference is the fitting procedure. The full zone models fit the bands over the full Brillouin zone, while in our approach we consider only the direct \emph{ab initio} matrix elements of $\bm{\pi} = \bm{p} + \bm{p}_{\rm SOC}$ without further manipulation.

If one needs a full zone model, we suggest using our results as the initial guess for the parameters used on a band-fitting algorithm. 
Moreover, since the fitted parameters must not deviate significantly from our \emph{ab initio} results, our calculated values provide an important benchmark for the fitting results.
Alternatively, it might be possible to develop multi-valley $\bm{k}\cdot\bm{p}$ models \cite{Persson2007, Marnetto2010, Berland2017} and extract its parameters directly from DFT matrix elements without numerical fitting procedures, but this is beyond the scope of this work.

\section{Conclusions}
\label{sec:conclusions}

We have implemented a numerical framework to calculate the $\bm{k}\cdot\bm{p}$ Kane and Luttinger parameters and optimal symmetry-adapted effective Hamiltonians directly from \emph{ab initio} wavefunctions. The code is mostly written in Python but also contains a patch to modify the Quantum ESPRESSO code, such that its \texttt{bands.x} post processing tool is used to calculate the matrix elements $\bm{P}_{m,n}=\bra{m}\bm{\pi}\ket{n}$, which is the central quantity in our methodology. Consequently, this first version works only with Quantum ESPRESSO. Equivalent calculations can be done in other DFT codes (e.g. VASP \cite{VASP}, Wien2k \cite{Wien2k}), but it requires further developments. The code is open source and it is available at Ref. \cite{DFT2KP}.

Here, we have illustrated the capabilities of our code applying it
to a series of relevant and well-known materials. The resulting effective
models yield band structures that match well the DFT data in the low
energy sector near the k point used for the wavefunction expansion.
Therefore, our code provides an \emph{ab initio} approach for the
$\bm{k}\cdot\bm{p}$ numerical parameters, which can be contrasted
with fitting methods
\cite{Mostofi2008, Mostofi2014, Paulo2016, BuongiornoNardelli2018, Gawarecki2022}, 
in which the numerical coefficients are obtained
by numerically minimizing the residue difference between the DFT and
model band structures over a selected range of the Brillouin zone.
These fitting procedures work well in general but require careful
verification if the fitted parameters are reasonable. In contrast,
our \emph{ab initio} approach is automatic and fully reliable. Nevertheless,
fitting procedures can improve the agreement between DFT and the model
band structures significantly. In this case, we suggest that our code
can be used (i) to generate the initial values for the fitting parameters,
and (ii) to verify if the fitted parameters show reasonable values.
One should expect that fitted parameters must not deviate much from
our \emph{ab initio} values.

Here we do not perform a thorough comparison of our numerical parameters
with experimental data. Typically, to obtain precise agreement with
experimental data, one needs to fix the gap issue by using either
hybrid functionals or GW calculations, which are beyond the scope
of this first version of the code. Instead, here we use only PBE functionals
\cite{GGA-PBE} for simplicity, which is reliable enough to validate our approach. Consequently, our numerical
parameters are limited by the precision of the DFT simulation, and
we would not expect remarkable agreement with experimental data for
most materials at this stage. Nevertheless, for novel materials, for
which there is no experimental data available, our code can be used
to generate reliable numerical parameters that can be improved later,
either in comparison with future experiments or by extending our
method to work with hybrid functionals or GW calculations.

As a final disclaimer, we would like to state that after developing
the first version of the code, we have found that Ref.~\cite{Jocic2020}
recently proposes an equivalent approach to build $\bm{k}\cdot\bm{p}$
models from DFT, but the authors do not provide an open-source code.
In any case, despite the similarities, the development of our code
was done independently from their proposal. In practice, the only
significant difference between the proposals is the approach to calculate
the transformation matrix $U$ (see Section \ref{subsec:symmopt}).
While the authors of Ref. \cite{Jocic2020} follow the method from
\cite{Mozrzymas2014}, here we propose a different method that is
more efficient for transformations involving reducible representations,
which is necessary when dealing with nearly degenerate bands of different
irreps (e.g., spinful graphene).
Additionally, after the initial submission of our paper, a new code \texttt{VASP2kp} \cite{vasp2kp}
was released with functionalities similar to ours, but designed for VASP \cite{VASP} instead of QE.

\begin{acknowledgments}
    This work was supported by the funding agencies CNPq, CAPES, and FAPEMIG. G.J.F.~acknowledges funding from the FAPEMIG grant PPM-00798-18).
    P.E.F.J.~acknowledges the financial support of the DFG SFB 1277 (Project-ID 314695032, projects B07 and B11) and SPP 2244 (Project No. 443416183).
    A.L.A.~acknowledges the financial support from FAPESP (grants 2022/08478-6 and 2023/12336-5).
    GFJ acknowledges useful discussions with P.~Giannozzi about PAW parameters on QE's pseudopotentials; H.~Zhao for useful discussions and the suggestion to use the \texttt{IrRep} and \texttt{qeirreps} \cite{Matsugatani2021} packages, and S.~S.~Tsirkin for discussions about the implementation of the \texttt{IrRep} package \cite{IrRep}.
\end{acknowledgments}

\newpage

\bibliography{qe2kp}

\begin{thebibliography}{136}
\bibitem [{VAS()}]{VASP2kpNote}%
  \BibitemOpen
  \href@noop {} {}\bibinfo {note} {{While our paper was in review, a new code
  \texttt{VASP2kp} \cite{vasp2kp}, similar to ours, was released.}}\BibitemShut
  {Stop}%
\bibitem [{C3v()}]{C3vDouble}%
  \BibitemOpen
  \href@noop {} {}\bibinfo {note} {{Equivalently, if one considers only
  in-plane symmetry operations the K point of spinful graphene transform as
  double group $C_{3v}$, and the basis functions are characterized by the
  representations $\bar{K}_4 \oplus (\bar{K}_5 \oplus \bar{K}_6)$ of this
  group. Here $\bar{K}_4$ is bidimensional (spinor irrep), while $\bar{K}_5
  \oplus \bar{K}_6$ are one-dimensinal irreps that combine to form Kramers
  partners.}}\BibitemShut {Stop}%
\bibitem [{Pso()}]{Psoc}%
  \BibitemOpen
  \href@noop {} {}\bibinfo {note} {{In this first version of the code we rely
  on the code \texttt{bands.x} from Quantum Expresso to calculate the matrix
  elements $\bm{P}_{m,n}$ with PAW/SOC corrections. However, as shown in
  Eq.~\eqref{eq:velocity}, this approach includes contribution from the
  mass-velocity term of the fine structure, which we assume to be negligible.
  For future improvements of the code, it would interesting to improve this
  calculation to verify if these mass-velocity contributions are indeed always
  negligible.}}\BibitemShut {Stop}%
\end{thebibliography}

\appendix

\newpage
\section{Mass-velocity corrections are negligible}
\label{sec:massvelocity}

Consider the full Hamiltonian with all fine structure corrections
as
\begin{align}
    H & =p^{2}+V(\bm{r})+H_{{\rm MV}}+H_{{\rm D}}+H_{{\rm SOC}},
    \\
    H_{{\rm MV}} & =-\frac{\alpha^{2}p^{4}}{4},
    \\
    H_{{\rm D}} & =\frac{\alpha^{2}}{8}\nabla^{2}V(\bm{r}),
    \\
    H_{{\rm SOC}} & =\frac{\alpha^{2}}{4}[\bm{\sigma}\times\nabla V(\bm{r})]\cdot\bm{p}.
\end{align}
Applying the Bloch theorem $\psi_{\bm{\kappa}}(\bm{r})=e^{i\bm{k}\cdot\bm{r}}\phi_{\bm{k}_{0},\bm{k}}(\bm{r})$
for $\bm{\kappa}=\bm{k}_{0}+\bm{k}$, the $\bm{k}\cdot\bm{p}$ Hamiltonian
becomes $H_{{\rm kp}}=H_{0}+k^{2}+H'$, where $H_{0}=p^{2}+V(\bm{r})+2\bm{k}_{0}\cdot\bm{\pi}+H_{{\rm SR}}$,
and $H_{{\rm SR}}$ contain the $\bm{k}=0$ contributions from $H_{{\rm MV}}+H_{{\rm D}}$,
as presented in the main text. The perturbation for finite $\bm{k}\neq0$
is $H'=2\bm{k}\cdot\bm{\pi}+H'_{{\rm MV}}$, where $H'_{{\rm MV}}$
contains the finite $\bm{k}$ contributions from the mass velocity
term, and it reads as
\begin{multline}
    H'_{{\rm MV}}=-\frac{\alpha^{2}}{4}\Big[4(\bm{k}\cdot\bm{p})p^{2}+4(\bm{k}\cdot\bm{p})^{2}
    \\
    +4(k)^{2}(\bm{k}\cdot\bm{p})+2k^{2}p^{2}+k^{4}\Big].
\end{multline}
These corrections are negligible for small $\bm{k}$, i.e. $|H'_{{\rm MV}}|\ll|2\bm{k}\cdot\bm{\pi}|$.
Notice that the SOC term in $2\bm{k}\cdot\bm{\pi}$ has two contributions,
one is of order $\sim|kp|$ and the other is $\sim|k\alpha^{2}|$.
In contrast, the contributions to $H'_{{\rm MV}}$ are $\sim|\alpha^{2}kp^{3}|$,
$\sim|\alpha^{2}k^{2}p^{2}|$, $\sim|\alpha^{2}k^{3}p|$, and $\sim|\alpha^{2}k^{4}|$.
Therefore, all terms in $H'_{{\rm MV}}$ are of higher order than
those in $2\bm{k}\cdot\bm{\pi}$, and we can safely assume $H'\approx2\bm{k}\cdot\bm{\pi}$.

\section{DFT parameters}
\label{sec:DFT}

The first principles calculations are performed using the density functional theory (DFT) \cite{dft1,dft2} within the generalized gradient approximation (GGA) for the exchange and correlation functional, employing the Perdew-Burke-Ernzerhof (PBE) parametrization \cite{GGA-PBE}. We employ the non-colinear spin-DFT formalism self-consistently with fully relativistic j-dependent ONCV (Optimized Norm-Conserving Vanderbilt) pseudopotential \cite{Hamann2013}. The Quantum ESPRESSO (QE) package \cite{Giannozzi2009,Giannozzi2017} was used, with a plane waves base configured with a given cut-off energy and the Brillouin zone sampled with several k-points (Monkhorst-Pack grid) so that the total energy converged within the meV scale (see Table \ref{tab:Convergence-criteria}). The ONCV pseudopotentials compatible with the Quantum ESPRESSO package are available in the repository \cite{repository-ONCV}. The vacuum space in two-dimensional materials was set to 15 Å. Atomic structures were optimized with a criterion that requires the force on each atom to be less than 0.01 eV/Å. Additional parameters used in our simulations including QE input and output files can be found in the \texttt{Examples} folder of the code repository \cite{DFT2KP}.

\begin{table}[t]
    \renewcommand*{\arraystretch}{1.5}
    \caption{\label{tab:Convergence-criteria} Criteria used for the convergence of the total energy: cut-off energy for the expansion in plane waves and the number of k-points taken for sampling the Brillouin zone using the Monkhorst-Pack technique.}
    \begin{ruledtabular}
    \begin{tabular}{l|r|r}
    Material & cut-off energy &  BZ sample \\
    \hline
    Graphene & 80 Ry & 12x12x1 \\
    GaAs & 100 Ry & 8x8x8 \\
    HgTe & 50 Ry & 8x8x8 \\
    CdTe & 60 Ry & 8x8x8 \\
    GaN & 100 Ry & 8x8x8 \\
    GaP & 150 Ry & 8x8x8 \\
    InP & 100 Ry & 7x7x7 \\
    PbSe & 100 Ry & 7x7x7 \\
    SnTe & 100 Ry & 8x8x8 \\    
    {${\rm MoS}_{2}$} & 100 Ry & 8x8x1 \\
    {${\rm Bi_{2}}{\rm Se}_{3}$} & 60 Ry & 7x7x7 \\
    {${\rm GaBiCl}_{2}$} & 100 Ry & 8x8x1 \\   
    \end{tabular}
    \end{ruledtabular}
\end{table}

\section{Effective Hamiltonians and coefficients}
\label{app:HandCs}

Here we present the large Hamiltonians and table of parameters for the materials presented in the main text. These correspond to the zincblende crystals for Fig.~\ref{fig:zincblende}, wurtzite crystals of Fig.~\ref{fig:wurtzite}, and rock-salt crystals of Fig.~\ref{fig:rocksalt}. For the other examples shown in Fig.~\ref{fig:OtherMaterials}, the corresponding Hamiltonians and numerical parameters can be seen in \texttt{Examples} folder in the code repository.

\begin{table}[H]
    \caption{\label{tab:ZBcs} Table of parameters for the zincblende materials, where the coefficients $c_{n}$ refer to the terms of $H_{{\rm ZB}}$ in the equation listed in Table \ref{tab:HZB}. The coefficient $c_{0}$ is negative for HgTe due to the $\Gamma_{6}$--$\Gamma_{8}$ band inversion.}
    \begin{ruledtabular}    
    \begin{tabular}{c|c|c|c}
    Zincblende & \text{GaAs} & \text{HgTe} & \text{CdTe}\\
    \hline
    $c_{0}$ (eV)           &  0.403    &  -1.16      &   0.36      \\
    $c_{1}$ (eV)           &  0.00011  &    2.23e-05 &   3.68e-05  \\
    $c_{2}$ (eV)           & -0.335    &  -0.773     &  -0.851     \\
    $c_{3}$ (eV~nm)        &  0.000486 &  -0.0117    &   0.00232   \\
    $c_{4}$ (eV~nm)        &  0.00268  &  -0.023     &   0.00499   \\
    $c_{5}$ (eV~nm)        & -0.635    &  -0.543     &   0.559     \\
    $c_{6}$ (eV~nm)        & -0.436    &    0.341    &   0.363     \\
    $c_{7}$ (eV~nm$^{2}$)  &  0.0293   &    0.0354   &   0.0347    \\
    $c_{8}$ (eV~nm$^{2}$)  & -0.0978   &  -0.0772    &  -0.0577    \\
    $c_{9}$ (eV~nm$^{2}$)  & -0.0437   &  -0.0339    &  -0.0262    \\
    $c_{10}$ (eV~nm$^{2}$) & -0.0321   &    0.0128   &  -0.0153    \\
    $c_{11}$ (eV~nm$^{2}$) & -0.0608   &  -0.0375    &  -0.0303    \\
    $c_{12}$ (eV~nm$^{2}$) & -0.000588 &  -0.0036    &  -0.000109  \\
    $c_{13}$ (eV~nm$^{2}$) &  0.0632   &    0.0558   &   0.0398    \\
    $c_{14}$ (eV~nm$^{2}$) &  0.0397   &  -0.0259    &   0.0231    \\
    $c_{15}$ (eV~nm$^{2}$) & -0.0362   &    0.0479   &   0.0361    \\
    $c_{16}$ (eV~nm$^{2}$) & -0.0275   &  -0.0349    &   0.0261    \\
    \end{tabular}
    \end{ruledtabular}
\end{table}

The numerical coefficients for the zincblende, wurtzite, and rock-salt materials are shown in Tables \ref{tab:ZBcs}, \ref{tab:WZcs}, and \ref{tab:RScs}, respectively. These correspond to the effective Hamiltonians shown in Tables \ref{tab:HZB}, \ref{tab:HWZ}, and \ref{tab:HRS}. In all cases we use $k_\pm = k_x \pm i k_y$, $k^2 = k_x^2 + k_y^2 + k_z^2$,
$k_\parallel^2 = k_x^2 + k_y^2$, $\hat{K} = k_x^2 - k_y^2$, which is also used in Appendix C of Ref.~\cite{Winkler2003}.


\begin{table}[H]
    \caption{\label{tab:RScs} Table of parameters for the rock-salt materials, where the coefficients $c_{n}$ refer to the terms of $H_{{\rm RS}}$ in the equation listed in Table \ref{tab:HRS}.}
    \begin{ruledtabular}    
    \begin{tabular}{c|c|c}
    Rock-salt & \text{PbSe} & \text{SnTe} \\
    \hline
    $c_{0}$ (eV)          &   0.235 &   0.125    \\
    $c_{1}$ (eV)          &   0.284 &   0.000141 \\
    $c_{2}$ (eV~nm)       &   0.168 &   0.193    \\
    $c_{3}$ (eV~nm)       &  -0.122 &  -0.111    \\
    $c_{4}$ (eV~nm$^{2}$) &  -0.134 &  -0.713    \\
    $c_{5}$ (eV~nm$^{2}$) &   0.223 &   0.214    \\
    $c_{6}$ (eV~nm$^{2}$) &   0.119 &   0.637    \\
    $c_{7}$ (eV~nm$^{2}$) &  -0.151 &  -0.158    \\
    \end{tabular}
    \end{ruledtabular}
\end{table}


\newpage
\begin{table}[H]
    \caption{\label{tab:WZcs} Table of parameters for the wurtzite materials, where the coefficients $c_{n}$ refer to the terms of $H_{{\rm WZ}}$ in the equation listed in Table \ref{tab:HWZ}.}
    \begin{ruledtabular}    
    \begin{tabular}{c|c|c|c}
    Wurtzite & \text{GaP} & \text{GaN} & \text{InP}\\
    \hline
    $c_{0}$ (eV)           &  1.75      &  1.76      &  0.457     \\
    $c_{1}$ (eV)           &  9.73e-06  & -1.16e-07  &  1.4e-07   \\
    $c_{2}$ (eV)           & -6.28e-06  & -5.09e-09  & -4.07e-06  \\
    $c_{3}$ (eV)           &  1.31      &  4.11      &  1.1       \\
    $c_{4}$ (eV)           & -0.208     & -0.0405    & -0.162     \\
    $c_{5}$ (eV)           &  4.7e-08   &  0.000658  &  6.89e-09  \\
    $c_{6}$ (eV)           & -0.0442    & -0.00602   & -0.0395    \\
    $c_{7}$ (eV)           &  7.82e-05  & -7.29e-05  &  8.11e-05  \\
    $c_{8}$ (eV~nm)        &  0.00448   &  0.00586   & -0.0112    \\
    $c_{9}$ (eV~nm)        &  0.00214   &  0.00075   &  0.0137    \\
    $c_{10}$ (eV~nm)       &  0.118     & -0.0733    & -0.184     \\
    $c_{11}$ (eV~nm)       &  0.455     & -0.372     & -0.392     \\
    $c_{12}$ (eV~nm)       & -0.472     & -0.381     &  0.436     \\
    $c_{13}$ (eV~nm)       & -0.00429   & -0.00428   & -0.0195    \\
    $c_{14}$ (eV~nm)       &  0.00811   &  0.0024    &  0.0223    \\
    $c_{15}$ (eV~nm)       &  0.0234    &  0.0128    &  0.0301    \\
    $c_{16}$ (eV~nm)       & -0.0268    &  0.0134    & -0.0428    \\
    $c_{17}$ (eV~nm)       & -0.0112    & -0.00416   & -0.0377    \\
    $c_{18}$ (eV~nm)       &  0.0055    & -0.00109   &  0.0202    \\
    $c_{19}$ (eV~nm)       &  0.801     & -0.568     & -0.616     \\
    $c_{20}$ (eV~nm)       &  0.214     & -0.116     & -0.298     \\
    $c_{21}$ (eV~nm)       & -0.00918   & -0.00423   & -0.0294    \\
    $c_{22}$ (eV~nm$^{2}$) &  0.0203    &  0.0266    &  0.0282    \\
    $c_{23}$ (eV~nm$^{2}$) &  0.0182    &  0.00155   & -0.0109    \\
    $c_{24}$ (eV~nm$^{2}$) &  0.00486   &  0.000225  & -0.00585   \\
    $c_{25}$ (eV~nm$^{2}$) & -2.32e-05  &  8.08e-05  & -0.000406  \\
    $c_{26}$ (eV~nm$^{2}$) &  0.273     &  0.128     &  0.264     \\
    $c_{27}$ (eV~nm$^{2}$) & -0.0267    & -0.0151    & -0.0262    \\
    $c_{28}$ (eV~nm$^{2}$) &  0.00735   &  0.00235   &  0.00881   \\
    $c_{29}$ (eV~nm$^{2}$) & -0.00672   &  0.00214   & -0.00733   \\
    $c_{30}$ (eV~nm$^{2}$) & -0.0558    & -0.0259    & -0.0411    \\
    $c_{31}$ (eV~nm$^{2}$) &  0.0285    & -0.0109    &  0.0178    \\
    $c_{32}$ (eV~nm$^{2}$) & -0.0581    & -0.0255    & -0.0433    \\
    $c_{33}$ (eV~nm$^{2}$) & -0.000342  &  0.00025   &  0.000387  \\
    $c_{34}$ (eV~nm$^{2}$) &  0.00537   & -0.00331   & -0.00541   \\
    $c_{35}$ (eV~nm$^{2}$) &  0.0223    & -0.0176    & -0.017     \\
    $c_{36}$ (eV~nm$^{2}$) &  0.0241    &  0.0197    & -0.0248    \\
    $c_{37}$ (eV~nm$^{2}$) &  0.00671   &  0.000335  & -0.00687   \\
    $c_{38}$ (eV~nm$^{2}$) &  0.0214    &  0.00371   & -0.00757   \\
    $c_{39}$ (eV~nm$^{2}$) & -0.0229    &  0.0034    &  0.0112    \\
    $c_{40}$ (eV~nm$^{2}$) &  0.00903   & -0.000739  &  0.00432   \\
    $c_{41}$ (eV~nm$^{2}$) & -0.00964   & -0.000377  & -0.0054    \\
    $c_{42}$ (eV~nm$^{2}$) &  0.00366   & -0.000194  &  0.00471   \\
    $c_{43}$ (eV~nm$^{2}$) & -7.74e-05  &  4.97e-06  &  0.000177  \\
    $c_{44}$ (eV~nm$^{2}$) &  0.0266    &  0.0241    &  0.0318    \\
    $c_{45}$ (eV~nm$^{2}$) & -0.0156    & -0.00776   &  0.00148   \\
    $c_{46}$ (eV~nm$^{2}$) & -0.00304   & -0.00173   & -0.0025    \\
    $c_{47}$ (eV~nm$^{2}$) &  0.0326    &  0.0175    &  0.031     \\
    $c_{48}$ (eV~nm$^{2}$) & -0.0661    & -0.0482    & -0.0636    \\
    $c_{49}$ (eV~nm$^{2}$) & -0.0107    & -0.00713   & -0.0205    \\
    $c_{50}$ (eV~nm$^{2}$) & -0.0334    & -0.0158    & -0.0379    \\
    $c_{51}$ (eV~nm$^{2}$) & -0.0294    & -0.0139    & -0.0242    \\
    \end{tabular}
    \end{ruledtabular}
\end{table}


\begin{turnpage}
\begin{table}
    \caption{\label{tab:HZB} Effective Hamiltonian for zincblende crystals considering the $8\times8$ extended Kane model.}
    \hrulefill
    \begin{multline*}
        H_{{\rm ZB}}=\left[\begin{matrix}c_{0}+c_{7}k^{2} & 0 & i\left(c_{15}k_{-}k_{z}-\frac{\sqrt{3}c_{5}k_{+}}{2}\right) & \frac{ic_{12}\cdot\left(2k_{z}^{2}-k_{\parallel}^{2}\right)}{2}\\
        0 & c_{0}+c_{7}k^{2} & \frac{ic_{12}\left(-2k_{z}^{2}+k_{\parallel}^{2}\right)}{2} & i\left(c_{15}k_{+}k_{z}+\frac{\sqrt{3}c_{5}k_{-}}{2}\right)\\
        i\left(-c_{15}k_{+}k_{z}+\frac{\sqrt{3}c_{5}k_{-}}{2}\right) & \frac{ic_{12}\cdot\left(2k_{z}^{2}-k_{\parallel}^{2}\right)}{2} & c_{1}+\frac{c_{8}\cdot\left(4k_{z}^{2}+k_{\parallel}^{2}\right)}{4}+\frac{3c_{9}k_{\parallel}^{2}}{4} & -\frac{\sqrt{3}c_{3}k_{-}}{2}\\
        \frac{ic_{12}\left(-2k_{z}^{2}+k_{\parallel}^{2}\right)}{2} & i\left(-c_{15}k_{-}k_{z}-\frac{\sqrt{3}c_{5}k_{+}}{2}\right) & -\frac{\sqrt{3}c_{3}k_{+}}{2} & c_{1}+\frac{c_{8}\cdot\left(4k_{z}^{2}+k_{\parallel}^{2}\right)}{4}+\frac{3c_{9}k_{\parallel}^{2}}{4}\\
        i\left(-\frac{\sqrt{3}c_{15}k_{-}k_{z}}{3}-\frac{c_{5}k_{+}}{2}\right) & -\frac{\sqrt{3}i\hat{K}c_{12}}{2}-\frac{2\sqrt{3}c_{15}k_{x}k_{y}}{3}-ic_{5}k_{z} & \frac{\sqrt{3}\hat{K}c_{8}}{4}-\frac{\sqrt{3}\hat{K}c_{9}}{4}+ic_{13}k_{x}k_{y}+c_{3}k_{z} & -c_{13}k_{-}k_{z}-\frac{c_{3}k_{+}}{2}\\
        \frac{\sqrt{3}i\hat{K}c_{12}}{2}-\frac{2\sqrt{3}c_{15}k_{x}k_{y}}{3}-ic_{5}k_{z} & i\left(-\frac{\sqrt{3}c_{15}k_{+}k_{z}}{3}+\frac{c_{5}k_{-}}{2}\right) & c_{13}k_{+}k_{z}-\frac{c_{3}k_{-}}{2} & \frac{\sqrt{3}\hat{K}c_{8}}{4}-\frac{\sqrt{3}\hat{K}c_{9}}{4}-ic_{13}k_{x}k_{y}-c_{3}k_{z}\\
        i\left(-c_{16}k_{-}k_{z}-c_{6}k_{+}\right) & c_{16}k_{x}k_{y}+ic_{6}k_{z} & \frac{\sqrt{3}\hat{K}c_{10}}{2}+2ic_{14}k_{x}k_{y}+c_{4}k_{z} & c_{14}k_{-}k_{z}-\frac{c_{4}k_{+}}{2}\\
        -c_{16}k_{x}k_{y}-ic_{6}k_{z} & i\left(c_{16}k_{+}k_{z}-c_{6}k_{-}\right) & c_{14}k_{+}k_{z}+\frac{c_{4}k_{-}}{2} & -\frac{\sqrt{3}\hat{K}c_{10}}{2}+2ic_{14}k_{x}k_{y}+c_{4}k_{z}
        \end{matrix}\right.\\\\\left.\begin{matrix}i\left(\frac{\sqrt{3}c_{15}k_{+}k_{z}}{3}+\frac{c_{5}k_{-}}{2}\right) & -\frac{\sqrt{3}i\hat{K}c_{12}}{2}-\frac{2\sqrt{3}c_{15}k_{x}k_{y}}{3}+ic_{5}k_{z} & i\left(c_{16}k_{+}k_{z}+c_{6}k_{-}\right) & -c_{16}k_{x}k_{y}+ic_{6}k_{z}\\
        \frac{\sqrt{3}i\hat{K}c_{12}}{2}-\frac{2\sqrt{3}c_{15}k_{x}k_{y}}{3}+ic_{5}k_{z} & i\left(\frac{\sqrt{3}c_{15}k_{-}k_{z}}{3}-\frac{c_{5}k_{+}}{2}\right) & c_{16}k_{x}k_{y}-ic_{6}k_{z} & i\left(-c_{16}k_{-}k_{z}+c_{6}k_{+}\right)\\
        \frac{\sqrt{3}\hat{K}c_{8}}{4}-\frac{\sqrt{3}\hat{K}c_{9}}{4}-ic_{13}k_{x}k_{y}+c_{3}k_{z} & c_{13}k_{-}k_{z}-\frac{c_{3}k_{+}}{2} & \frac{\sqrt{3}\hat{K}c_{10}}{2}-2ic_{14}k_{x}k_{y}+c_{4}k_{z} & c_{14}k_{-}k_{z}+\frac{c_{4}k_{+}}{2}\\
        -c_{13}k_{+}k_{z}-\frac{c_{3}k_{-}}{2} & \frac{\sqrt{3}\hat{K}c_{8}}{4}-\frac{\sqrt{3}\hat{K}c_{9}}{4}+ic_{13}k_{x}k_{y}-c_{3}k_{z} & c_{14}k_{+}k_{z}-\frac{c_{4}k_{-}}{2} & -\frac{\sqrt{3}\hat{K}c_{10}}{2}-2ic_{14}k_{x}k_{y}+c_{4}k_{z}\\
        c_{1}+\frac{3c_{8}k_{\parallel}^{2}}{4}+\frac{c_{9}\cdot\left(4k_{z}^{2}+k_{\parallel}^{2}\right)}{4} & \frac{\sqrt{3}c_{3}k_{-}}{2} & \frac{c_{10}\cdot\left(2k_{z}^{2}-k_{\parallel}^{2}\right)}{2} & \sqrt{3}\left(-c_{14}k_{+}k_{z}+\frac{c_{4}k_{-}}{2}\right)\\
        \frac{\sqrt{3}c_{3}k_{+}}{2} & c_{1}+\frac{3c_{8}k_{\parallel}^{2}}{4}+\frac{c_{9}\cdot\left(4k_{z}^{2}+k_{\parallel}^{2}\right)}{4} & \sqrt{3}\left(-c_{14}k_{-}k_{z}-\frac{c_{4}k_{+}}{2}\right) & \frac{c_{10}\left(-2k_{z}^{2}+k_{\parallel}^{2}\right)}{2}\\
        \frac{c_{10}\cdot\left(2k_{z}^{2}-k_{\parallel}^{2}\right)}{2} & \sqrt{3}\left(-c_{14}k_{+}k_{z}-\frac{c_{4}k_{-}}{2}\right) & c_{11}k^{2}+c_{2} & 0\\
        \sqrt{3}\left(-c_{14}k_{-}k_{z}+\frac{c_{4}k_{+}}{2}\right) & \frac{c_{10}\left(-2k_{z}^{2}+k_{\parallel}^{2}\right)}{2} & 0 & c_{11}k^{2}+c_{2}
        \end{matrix}\right]
    \end{multline*}
    \hrulefill
\end{table}
\end{turnpage}


\begin{turnpage}
\begin{table}
    \caption{\label{tab:HWZ} Effective Hamiltonian for wurtzite crystals considering the $10\times10$ model with two conduction bands.}
    \hrulefill
    \begin{multline*}
        H_{{\rm WZ}}=
        \left[\begin{matrix}
 c_{0} + c_{22} k_{\parallel}^2 + c_{44} k_{z}^{2}  &  i c_{9} k_{-}  &  0  &  c_{33} \left(i k_{x}^{2} - 2 k_{x} k_{y} - i k_{y}^{2}\right)  &  c_{1} + i c_{19} k_{z} + c_{23} k_{\parallel}^2 + c_{45} k_{z}^{2} 
 \\ 
 - i c_{9} k_{+}  &  c_{0} + c_{22} k_{\parallel}^2 + c_{44} k_{z}^{2}  &  c_{33} \left(i k_{x}^{2} + 2 k_{x} k_{y} - i k_{y}^{2}\right)  &  0  &  - k_{+} \left(i c_{10} + c_{37} k_{z}\right) 
 \\ 
 0  &  c_{33} \left(- i k_{x}^{2} + 2 k_{x} k_{y} + i k_{y}^{2}\right)  &  c_{26} k_{\parallel}^2 + c_{3} + c_{47} k_{z}^{2}  &  i c_{13} k_{-}  &  0 
 \\ 
 c_{33} \left(- i k_{x}^{2} - 2 k_{x} k_{y} + i k_{y}^{2}\right)  &  0  &  - i c_{13} k_{+}  &  c_{26} k_{\parallel}^2 + c_{3} + c_{47} k_{z}^{2}  &  c_{34} \left(i k_{x}^{2} + 2 k_{x} k_{y} - i k_{y}^{2}\right) 
 \\ 
 c_{1} - i c_{19} k_{z} + c_{23} k_{\parallel}^2 + c_{45} k_{z}^{2}  &  k_{-} \left(i c_{10} - c_{37} k_{z}\right)  &  0  &  c_{34} \left(- i k_{x}^{2} + 2 k_{x} k_{y} + i k_{y}^{2}\right)  &  c_{27} k_{\parallel}^2 + c_{4} + c_{48} k_{z}^{2} 
 \\ 
 k_{+} \left(- i c_{10} + c_{37} k_{z}\right)  &  c_{1} - i c_{19} k_{z} + c_{23} k_{\parallel}^2 + c_{45} k_{z}^{2}  &  c_{34} \left(- i k_{x}^{2} - 2 k_{x} k_{y} + i k_{y}^{2}\right)  &  0  &  - i c_{14} k_{+} 
 \\ 
 k_{+} \left(- i c_{11} + c_{38} k_{z}\right)  &  - c_{2} + i c_{20} k_{z} - c_{24} k_{\parallel}^2 - c_{46} k_{z}^{2}  &  c_{35} \left(- i k_{x}^{2} - 2 k_{x} k_{y} + i k_{y}^{2}\right)  &  0  &  k_{+} \left(- i c_{15} + c_{40} k_{z}\right) 
 \\ 
 c_{2} - i c_{20} k_{z} + c_{24} k_{\parallel}^2 + c_{46} k_{z}^{2}  &  k_{-} \left(- i c_{11} + c_{38} k_{z}\right)  &  0  &  c_{35} \left(i k_{x}^{2} - 2 k_{x} k_{y} - i k_{y}^{2}\right)  &  - i c_{21} k_{z} + c_{28} k_{\parallel}^2 + c_{49} k_{z}^{2} + c_{5} 
 \\ 
 k_{-} \left(- i c_{12} + c_{39} k_{z}\right)  &  c_{25} \left(- \hat{K} + 2 i k_{x} k_{y}\right)  &  c_{36} \left(- i k_{x}^{2} + 2 k_{x} k_{y} + i k_{y}^{2}\right)  &  k_{+} \left(- i c_{43} k_{z} + c_{8}\right)  &  k_{-} \left(- i c_{16} + c_{41} k_{z}\right) 
 \\ 
 c_{25} \left(\hat{K} + 2 i k_{x} k_{y}\right)  &  k_{+} \left(- i c_{12} + c_{39} k_{z}\right)  &  k_{-} \left(- i c_{43} k_{z} + c_{8}\right)  &  c_{36} \left(i k_{x}^{2} + 2 k_{x} k_{y} - i k_{y}^{2}\right)  &  c_{29} \left(\hat{K} + 2 i k_{x} k_{y}\right) 
\end{matrix}\right.
\\
\left.\begin{matrix}
 k_{-} \left(i c_{10} + c_{37} k_{z}\right)  &  k_{-} \left(i c_{11} + c_{38} k_{z}\right)  &  c_{2} + i c_{20} k_{z} + c_{24} k_{\parallel}^2 + c_{46} k_{z}^{2}  &  k_{+} \left(i c_{12} + c_{39} k_{z}\right)  &  c_{25} \left(\hat{K} - 2 i k_{x} k_{y}\right)
 \\ 
 c_{1} + i c_{19} k_{z} + c_{23} k_{\parallel}^2 + c_{45} k_{z}^{2}  &  - c_{2} - i c_{20} k_{z} - c_{24} k_{\parallel}^2 - c_{46} k_{z}^{2}  &  k_{+} \left(i c_{11} + c_{38} k_{z}\right)  &  - c_{25} \left(\hat{K} + 2 i k_{x} k_{y}\right)  &  k_{-} \left(i c_{12} + c_{39} k_{z}\right)
 \\ 
 c_{34} \left(i k_{x}^{2} - 2 k_{x} k_{y} - i k_{y}^{2}\right)  &  c_{35} \left(i k_{x}^{2} - 2 k_{x} k_{y} - i k_{y}^{2}\right)  &  0  &  c_{36} \left(i k_{x}^{2} + 2 k_{x} k_{y} - i k_{y}^{2}\right)  &  k_{+} \left(i c_{43} k_{z} + c_{8}\right)
 \\ 
 0  &  0  &  c_{35} \left(- i k_{x}^{2} - 2 k_{x} k_{y} + i k_{y}^{2}\right)  &  k_{-} \left(i c_{43} k_{z} + c_{8}\right)  &  c_{36} \left(- i k_{x}^{2} + 2 k_{x} k_{y} + i k_{y}^{2}\right)
 \\ 
 i c_{14} k_{-}  &  k_{-} \left(i c_{15} + c_{40} k_{z}\right)  &  i c_{21} k_{z} + c_{28} k_{\parallel}^2 + c_{49} k_{z}^{2} + c_{5}  &  k_{+} \left(i c_{16} + c_{41} k_{z}\right)  &  c_{29} \left(\hat{K} - 2 i k_{x} k_{y}\right)
 \\ 
 c_{27} k_{\parallel}^2 + c_{4} + c_{48} k_{z}^{2}  &  - i c_{21} k_{z} - c_{28} k_{\parallel}^2 - c_{49} k_{z}^{2} - c_{5}  &  k_{+} \left(i c_{15} + c_{40} k_{z}\right)  &  - c_{29} \left(\hat{K} + 2 i k_{x} k_{y}\right)  &  k_{-} \left(i c_{16} + c_{41} k_{z}\right)
 \\ 
 i c_{21} k_{z} - c_{28} k_{\parallel}^2 - c_{49} k_{z}^{2} - c_{5} &  c_{30} k_{\parallel}^2 + c_{50} k_{z}^{2} + c_{6}  &  i c_{17} k_{+}  &  c_{31} \left(\hat{K} + 2 i k_{x} k_{y}\right)  &  k_{-} \left(i c_{18} + c_{42} k_{z}\right)
 \\ 
 k_{-} \left(- i c_{15} + c_{40} k_{z}\right)  &  - i c_{17} k_{-}  &  c_{30} k_{\parallel}^2 + c_{50} k_{z}^{2} + c_{6}  &  - k_{+} \left(i c_{18} + c_{42} k_{z}\right) &  c_{31} \left(\hat{K} - 2 i k_{x} k_{y}\right)
 \\ 
 c_{29} \left(- \hat{K} + 2 i k_{x} k_{y}\right)  &  c_{31} \left(\hat{K} - 2 i k_{x} k_{y}\right)  &  k_{-} \left(i c_{18} - c_{42} k_{z}\right)  &  c_{32} k_{\parallel}^2 + c_{51} k_{z}^{2} + c_{7}  &  0
 \\ 
 k_{+} \left(- i c_{16} + c_{41} k_{z}\right)  &  k_{+} \left(- i c_{18} + c_{42} k_{z}\right)  &  c_{31} \left(\hat{K} + 2 i k_{x} k_{y}\right)  &  0  &  c_{32} k_{\parallel}^2 + c_{51} k_{z}^{2} + c_{7}
\end{matrix}\right]
    \end{multline*}
    \hrulefill
\end{table}
\end{turnpage}


\begin{table*}
    \caption{\label{tab:HRS} Effective Hamiltonian for rock-salt crystals considering the $4\times4$ model composed by
the $L_{6}^{\pm}$ irreps of $D_{3D}$.}
    \hrulefill
    \begin{multline*}
        H_{{\rm RS}}=
        \left[\begin{matrix}
            c_{0}+c_{4}k^{2}+c_{6}\left(k_{x}k_{y}+k_{x}k_{z}+k_{y}k_{z}\right) & 0
            \\
            0 & c_{0}+c_{4}k^{2}+c_{6}\left(k_{x}k_{y}+k_{x}k_{z}+k_{y}k_{z}\right)
            \\
            -c_{2}\left(k_{x}-k_{y}\right)-ic_{3}\left(k_{x}+k_{y}+k_{z}\right) & c_{2}\left(-ik_{-}+k_{z}\left(1+i\right)\right)
            \\
            c_{2}\left(ik_{+}+k_{z}\left(1-i\right)\right) & c_{2}\left(k_{x}-k_{y}\right)-ic_{3}\left(k_{x}+k_{y}+k_{z}\right)
        \end{matrix}\right.
        \\\\
        \left.\begin{matrix}
            -c_{2}\left(k_{x}-k_{y}\right)+ic_{3}\left(k_{x}+k_{y}+k_{z}\right) & c_{2}\left(-ik_{-}+k_{z}\left(1+i\right)\right)
            \\
            c_{2}\left(ik_{+}+k_{z}\left(1-i\right)\right) & c_{2}\left(k_{x}-k_{y}\right)+ic_{3}\left(k_{x}+k_{y}+k_{z}\right)
            \\
            c_{1}+c_{5}k^{2}+c_{7}\left(k_{x}k_{y}+k_{x}k_{z}+k_{y}k_{z}\right) & 0
            \\
            0 & c_{1}+c_{5}k^{2}+c_{7}\left(k_{x}k_{y}+k_{x}k_{z}+k_{y}k_{z}\right)
        \end{matrix}\right]
    \end{multline*}
    \hrulefill
\end{table*}

\end{document}